\documentclass[iop,revtex4]{emulateapj}
\usepackage{apjfonts}
\usepackage{amssymb,appendix,multirow,morefloats,amsmath}
\usepackage[figuresright]{rotating}
\usepackage{threeparttable}
\usepackage{longtable}
\usepackage{subfigure}
\usepackage{epstopdf}
\usepackage{graphicx}
\usepackage{pdflscape}
\usepackage{chngpage}

\newcommand\T{\rule{0pt}{2.6ex}}       % Top strut
\newcommand\B{\rule[-1.2ex]{0pt}{0pt}} % Bottom strut

\begin{document}

\title{An \textit{XMM-Newton} and \textit{NuSTAR} study of IGR J18214-1318: a non-pulsating high-mass X-ray binary with a neutron star}

\shorttitle{IGR J18214-1318}
 
\shortauthors{Fornasini et al.}
\slugcomment{Accepted to ApJ: April 27, 2017}

\author{Francesca M. Fornasini\altaffilmark{1,2,3}, John A. Tomsick\altaffilmark{2}, Matteo Bachetti\altaffilmark{4}, Roman A. Krivonos\altaffilmark{2,5}, Felix F{\"u}rst\altaffilmark{6}, Lorenzo Natalucci\altaffilmark{7}, Katja Pottschmidt\altaffilmark{8,9}, and J{\"o}rn Wilms\altaffilmark{10}}

\altaffiltext{1}{Astronomy Department, University of California, 601 Campbell Hall, Berkeley, CA 94720, USA (e-mail: f.fornasini@berkeley.edu)}
\altaffiltext{2}{Space Sciences Laboratory, 7 Gauss Way, University of California, Berkeley, CA 94720, USA}
\altaffiltext{3}{Harvard-Smithsonian Center for Astrophysics, 60 Garden St., Cambridge, MA 02138, USA}
\altaffiltext{4}{INAF/Osservatorio Astronomico di Cagliari, via della Scienza 5, I-09047 Selargius (CA), Italy}
\altaffiltext{5}{Space Research Institute, Russian Academy of Sciences, Profsoyuznaya 84/32, 117997 Moscow, Russia}
\altaffiltext{6}{Cahill Center for Astronomy and Astrophysics, California Institute of Technology, Pasadena, CA 91125, USA}
\altaffiltext{7}{Istituto di Astrofisica e Planetologia Spaziali, INAF, Via Fosso del Cavaliere 100, I-00133 Roma, Italy}
\altaffiltext{8}{Department of Physics and Center for Space Science and Technology, University of Maryland Baltimore County, Baltimore, MD 21250, USA}
\altaffiltext{9}{3 CRESST and NASA Goddard Space Flight Center, Astrophysics Science Division, Code 661, Greenbelt, MD 20771, USA}
\altaffiltext{10}{Dr. Karl-Remeis-Sternwarte and ECAP, Sternwartstr. 7, D-96049 Bamberg, Germany}

\begin{abstract}
\noindent IGR J18214-1318, a Galactic source discovered by the \textit{International Gamma-Ray Astrophysics Laboratory}, is a high-mass X-ray binary (HMXB) with a supergiant O-type stellar donor.  We report on the \textit{XMM-Newton} and \textit{NuSTAR} observations that were undertaken to determine the nature of the compact object in this system.  This source exhibits high levels of aperiodic variability, but no periodic pulsations are detected with a 90\% confidence upper limit of 2\% fractional rms between 0.00003--88 Hz, a frequency range that includes the typical pulse periods of neutron stars (NSs) in HMXBs (0.1--10$^3$ s).  Although the lack of pulsations prevents us from definitively identifying the compact object in IGR J18214-1318, the presence of an exponential cutoff with e-folding energy $\lesssim30$ keV in its 0.3-79 keV spectrum strongly suggests that the compact object is an NS.  The X-ray spectrum also shows a Fe K$\alpha$ emission line and a soft excess, which can be accounted for by either a partial-covering absorber with $N_{\mathrm{H}}\approx10^{23}$ cm$^{-2}$ which could be due to the inhomogeneous supergiant wind, or a blackbody component with $kT=1.74^{+0.04}_{-0.05}$ keV and $R_{BB}\approx0.3$ km, which may originate from NS hot spots.  Although neither explanation for the soft excess can be excluded, the former is more consistent with the properties observed in other supergiant HMXBs.  We compare IGR J18214-1318 to other HMXBs that lack pulsations or have long pulsation periods beyond the range covered by our observations.  

\end{abstract}

\keywords{stars: individual (IGR J18214-1318) -- X-rays: binaries -- stars: neutron -- stars: black holes}

\section{Introduction}
\label{sec:intro}

High-mass X-ray binaries (HMXBs) inform our understanding of the evolution of massive stars, which is still subject to significant uncertainties \citep{smith14}.  Studying the accreting neutron stars (NSs) and black holes (BHs) in these systems offers a special tool to probe the strength and clumping of the stellar winds of their massive companions.  Moreover, comparing the properties of HMXB populations to predictions of population synthesis models 
%(i.e. distributions of their orbital periods, compact object masses, and donor spectral types) 
helps constrain theoretical models of stellar mass loss, mass transfer episodes in massive binaries, and the natal kicks received by compact objects during supernova explosions (e.g.,
\citealt{negueruela08}; \citealt{linden10}; \citealt{fragos13}; \citealt{grudzinska15}).  
%Since HMXBs are the likely progenitors of many of the double compact binaries which may merge and produce gravitational waves \citep{postnov14}, such as the BH-BH merger recently detected by LIGO \citep{abbott16}, studies of HMXB populations naturally complement the new field of gravitational wave astronomy and will be useful in determining the implications of gravitational wave sources for stellar evolutionary models.  
Constraining the ratio of NS to BH HMXBs and whether this ratio varies with binary properties (e.g., donor spectral type and metallicity) can shed light on the net mass loss experienced by a high-mass star due to its stellar wind, binary interactions, and supernova explosion (\citealt{dray06}; \citealt{muno07}; \citealt{belczynski09}) as well as improve our estimates of the relative fractions of different double compact binaries expected to descend from HMXBs \citep{postnov14}.  \par
HMXBs hosting BHs exhibit different spectral properties from those hosting NSs.  NS HMXBs typically have hard power-law spectra with exponential cutoffs, with e-folding energies typically $\lesssim30$ keV \citep{coburn02}, whereas BH HMXBs exhibit power-law cutoffs around 50--100 keV in their hard states and $\Gamma\sim2$ power-law tails extending to MeV energies in their soft states (\citealt{grove98}; \citealt{zdziarski00}).  Thus, the presence of an exponential cutoff below 20 keV in the X-ray spectrum of an HMXB is a strong indication that it harbors an NS; however, only the detection of X-ray pulsations or cyclotron line features constitute definitive proof of the presence of an NS  Most X-ray pulsars in HMXBs have spin periods between $\sim0.1$ and $\sim10^3$ s (\citealt{corbet86}; \citealt{chaty13}), although a couple of longer-period pulsars have been discovered (\citealt{reig09}; \citealt{corbet99}).  In addition to NS X-ray pulsations, the X-ray light curves of some HMXBs can exhibit orbital or superorbital modulations with typical periods of a few hours to a few hundred days (\citealt{corbet06}; \citealt{corbet13}).  The cyclotron lines that have been observed in some HMXB spectra have energies between 10 and 80 keV, corresponding to magnetic field strengths of a few $10^{12}$ G (\citealt{coburn02}; \citealt{pottschmidt05}; \citealt{caballero07}; \citealt{doroshenko10}; \citealt{caballero12}; \citealt{tsygankov12}; \citealt{fuerst14}; \citealt{tendulkar14}; \citealt{yamamoto14}; \citealt{bellm14}).  An HMXB with an unbroken power-law spectrum extending beyond 50 keV that does not show X-ray pulsations can be considered a BH candidate, but confirming the BH nature of the compact object requires a dynamical mass measurement in excess of 2--3 $M_{\odot}$, the maximum theoretically expected NS mass \citep{lattimer12}; however, such measurements can be challenging to obtain.  \par
Since its launch in 2002, the \textit{International Gamma-Ray Astrophysics Laboratory (INTEGRAL)} has discovered a large number of sources that were given "IGR" source names, including 40 new HMXBs in the 4th IBIS/ISGRI catalog \citep{bird16}.  X-ray pulsations have only been detected from about a quarter of the IGR HMXBs, and the nature of the compact object in the remaining systems is undetermined since many of them are relatively recent discoveries and have not yet been well-studied.  Identifying the nature of compact objects in IGR HMXBs is of special interest since they are demographically different from the population of HMXBs discovered prior to \textit{INTEGRAL}, consisting of roughly equal numbers of HMXBs with Be and supergiant (Sg) stellar companions, whereas Be HMXBs dominate the pre-\textit{INTEGRAL} population \citep{walter15}.  Many of the IGR Sg HMXBs have unusual properties, exhibiting high levels of obscuration ($N_{\mathrm{H}}\sim10^{23}-10^{24}$ cm$^{-2}$) local to the source (\citealt{walter06}; \citealt{chaty08}) or extreme flaring behavior characterized by hard X-ray flux variations of several orders of magnitude on timescales of a few hours (\citealt{negueruela06}; \citealt{sguera06}).  The discovery of these Sg HMXBs was made possible by the greater sensitivity and higher cadence of the \textit{INTEGRAL} Galactic Plane survey compared to previous hard X-ray missions.  \par
A few of the flaring IGR Sg HMXBs, known as supergiant fast X-ray transients (SFXTs), are known to host NSs based on the detection of X-ray pulsations \citep{romano14} or cyclotron lines \citep{bhalerao15}, and several of the models proposed to explain SFXT behavior depend on the presence of an NS magnetosphere (the propeller effect, \citealt{grebenev07}; magnetic gating, \citealt{bozzo08}; quasi-spherical settling accretion of hot plasma shells, \citealt{shakura14}), suggesting that all SFXTs may host NSs.  However, the nature of compact objects in many IGR HMXBs that are not SFXTs remains unknown, and as several Sg HMXBs are known to harbor BHs (e.g., Cyg X-1, M33 X-7, LMC X-1, LMC X-3) compared to only one known Be-BH binary \citep{casares14}, IGR Sg HMXBs constitute a particularly promising group to search for BHs.  \par
IGR J18214-1318 is one of the Sg HMXBs lacking a clear compact object identification.  This source was first reported in the second IBIS/ISGRI catalog \citep{bird06} and detected consistently by \textit{INTEGRAL} with a flux of 1--2 mCrab in the 20-40 keV band (\citealt{krivonos12}; \citealt{bird16}).  The source was localized with arcsecond precision to R.A. = $18^{\mathrm{h}}21^{\mathrm{m}}19.76^{\mathrm{s}}$, decl. = $-13^{\circ}18^{\prime}38.9^{\prime\prime}$ through a \textit{Chandra} observation \citep{tomsick08}.  The localization of this source permitted its association with an optical counterpart, which is a high-mass star of most likely spectral type O9 I \citep{butler09}, thus securing the identification of IGR J18214-1318 as a Sg HMXB.  Its \textit{Chandra} spectrum is well-fit by an absorbed power law with $N_{\mathrm{H}} = 1.2\pm0.3 \times 10^{23}$ cm$^{-2}$ and $\Gamma=0.7^{+0.6}_{-0.5}$.  Later \textit{Swift} observations of this source measured a similar photon index ($\Gamma=0.4\pm0.2$) but a much lower absorbing column density of $N_{\mathrm{H}} = 3.5^{+0.8}_{-0.5} \times 10^{22}$ cm$^{-2}$, which is consistent with the Galactic $N_{\mathrm{H}}$ integrated along the line of sight (3.1$\times10^{22}$ cm$^{-2}$)\footnotemark\footnotetext{The Galactic $N_{\mathrm{H}}$ along the line of sight to IGR J18214-1318 is calculated as the sum of the HI contribution measured from the Leiden/Argentine/Bonn (LAB) survey of HI \citep{kalberla05} and the H$_2$ contribution estimated from the MWA CO survey \citep{bronfman89}.}.  Although the \textit{Chandra} data suffered from photon pile-up, since the photon indices derived by \textit{Chandra} and \textit{Swift} are so similar, it is unlikely that the large difference in the derived $N_{\mathrm{H}}$ values is simply a result of the photon pile-up in \textit{Chandra}.  Thus, the large variability of $N_{\mathrm{H}}$ is likely real and associated with material local to the source; similar $N_{\mathrm{H}}$ variations have been seen in other Sg HMXBs (e.g, IGR J19140+0951; \citealt{prat08}).  The hard power-law index measured in the soft X-ray band suggests that the compact object in IGR J18214-1318 is more likely to be an NS than a BH, but it does not constitute strong or definitive evidence.  \par
Therefore, we observed IGR J18214-1318 with the \textit{Nuclear Spectroscopic Telescope Array (NuSTAR)} and \textit{XMM-Newton} (\S\ref{sec:obs}) to better constrain the nature of the compact object in this HMXB.  \textit{NuSTAR} and \textit{XMM-Newton} are ideally suited for this study, because their instruments have the fast temporal resolution required to search for X-ray pulsations and their combined broadband X-ray spectral coverage from 0.3-79 keV permits the measurement of cyclotron lines and hard X-ray cutoffs, which may be present in the HMXB power-law spectrum.  Spectral analysis of this data, resulting in the detection of a high-energy cutoff, is described in \S\ref{sec:spectral}, and timing analysis ruling out the existence of a pulse period $\lesssim1$ hr is presented in \S\ref{sec:timing}.  In \S\ref{sec:discussion}, we discuss IGR J18214-1318 in the context of other non-pulsating and long pulse period HMXBs.

%INTEGRAL catalog:
%9 candidate HMXBs (8 IGR)
%67 non-IGR, 40 IGR HMXBs > 10 XPs + additionally 8 SFXTs (probably NS) + 7 Be without confirmed 

\begin{table*}
\centering
\footnotesize
\caption{Observations of IGR J18214-1318}
\begin{threeparttable}
\begin{tabular}{cccccc} \hline \hline
\T Telescope & Observation ID & Start Time & Duration & Instrument/ & Exposure \\ \cline{3-4}
\T & & (UTC) & (ks) & Detector & (ks) \\
\B (1) & (2) & (3) & (4) & (5) & (6) \\
\hline
\T \multirow{2}{*}{\textit{NuSTAR}} & \multirow{2}{*}{3000114002} & \multirow{2}{*}{2014 Sep 18 01:16:07} & \multirow{2}{*}{49.8} & FPMA & 26.0 \\
&&&&FPMB& 25.8 \\
\hline
\T \multirow{3}{*}{\textit{XMM-Newton}} & \multirow{3}{*}{0741470201} & \multirow{3}{*}{2014 Sep 18 02:34:26} & \multirow{3}{*}{26.9} & EPIC pn & 18.5 \\
&&&&MOS1& 25.9\\
&&&&MOS2& 25.8\\
\hline
\end{tabular}
\begin{tablenotes}[flushleft]
\item Notes: (6) Exposure does not include dead time.
\end{tablenotes}
\end{threeparttable}
\label{tab:obs}
\end{table*}

\section{Observations and Data Reduction}
\label{sec:obs}

\textit{NuSTAR} and \textit{XMM-Newton} observed IGR J18214-1318 on 2014 September 18.  Observation details are provided in Table \ref{tab:obs}.  The duration of the \textit{NuSTAR} observation is about twice as long as that of \textit{XMM-Newton} because Earth occultations reduce the effective exposure of \textit{NuSTAR}.  The exposure times of the two \textit{NuSTAR} focal plane modules (FPM) and the two \textit{XMM} Metal Oxide Semi-conductor (MOS) CCD arrays of the European Photon Imaging Camera (EPIC) are roughly equal (26 ks), while the exposure time of the EPIC pn CCD is significantly lower (18.5 ks) due to its higher dead-time fraction.  

\subsection{\textit{NuSTAR}}

\begin{figure}[b]
\vspace{0.1in}
\includegraphics[width=0.48\textwidth]{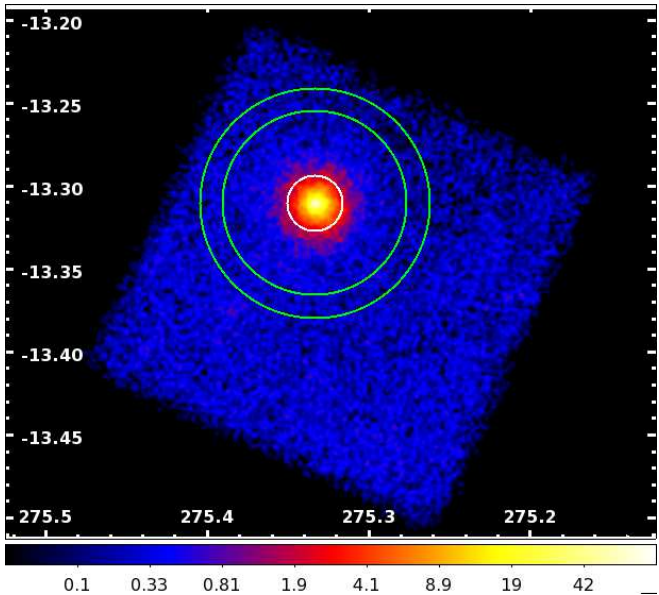}
\caption{The \textit{NuSTAR} FPMA observation in the 3--79 keV band smoothed by a Gaussian kernel with a 3 pixel radius.  The colorbar displays the counts per pixel on a logarithmic scale.  The source region is represented by a white circle, and the background region is represented by a green annulus.}
\label{fig:nustarobs}
\end{figure}

\textit{NuSTAR} is the first hard X-ray focusing telescope in space, providing 58$^{\prime\prime}$ half-power diameter angular resolution \citep{harrison13}.  The \textit{NuSTAR} FPMs cover the 3-79 keV band with moderate energy resolution (0.4 keV energy at 6 keV) and operate with 0.1 ms temporal resolution \citep{harrison13}.  We processed the data from the two \textit{NuSTAR} instruments, FPMA and FPMB, with the NuSTARDAS pipeline software v1.4.1, the 20150612 version of the \textit{NuSTAR} Calibration Database (CALDB), and High Energy Astrophysics Software (HEASOFT) version 6.16.    \par
Cleaned event lists were produced with the routine \texttt{nupipeline}.  We used \texttt{nuproducts} to extract spectra, including response matrix files (RMFs) and ancillary response files (ARFs), and light curves, applying barycenter and dead-time corrections.  In order to choose the aperture region sizes, we measured the surface brightness of the profile of the source by measuring the average count rate per pixel in concentric annuli centered on the source.  We found that the profile substantially flattens at a distance of 150$^{\prime\prime}$ from the source, so to guarantee a clean background measurement, we defined the background region as an annulus centered on the source with an inner radius of 200$^{\prime\prime}$ and an outer radius of 250$^{\prime\prime}$.  Within 60$^{\prime\prime}$ of the source, the count rate per pixel is at least 10 times higher than in the background region, so we defined the source aperture region as a circle with a 60$^{\prime\prime}$ radius (corresponding to the PSF encircled energy fraction of 75\%) to limit background contamination while still ensuring good photon statistics ($\sim$30,000 counts in FPMA and FPMB combined).  Figure \ref{fig:nustarobs} displays the \textit{NuSTAR} FPMA observation with the source and background regions. We checked the light curves from the background regions for significant count rate variations, but did not find any significant background variability in the 3--79, 3--12, or 12--30 keV bands.  The 3-30 keV dead-time-corrected source count rate is 1.2 counts s$^{-1}$ for FPMA and FPMB combined.

%angular resolution (half-power diameter) of 58, 0.1 msec, 12'x12', 0.4 keV at 6 keV, 0.9 keV at 60 keV

\subsection{\textit{XMM-Newton}}

We made use of data from the EPIC pn and MOS instruments on \textit{XMM-Newton}, which provide 15$^{\prime\prime}$ HPD angular resolution at soft X-ray energies. Observations were performed in small window mode with a medium filter.  The EPIC pn instrument covers the 0.3--12 keV band with an energy resolution of 150 eV at 6 keV and, in the small window mode, it has 5.7 ms time resolution \citep{struder01}.  The MOS cameras provide similar energy resolution in the 0.3--10 keV band but have poorer timing resolution of 0.3 s in the small window mode \citep{turner01}.  

We processed the {\em XMM-Newton} data with Science Analysis Software (SAS) v13.5.0, making images, spectra, and light curves for EPIC pn, MOS1, and MOS2.  To look for contamination from proton flares, we made EPIC pn and MOS light curves in the 10--12\,keV bandpass, but we did not find any significant flares.  For all three instruments, we made new event lists using the standard filtering criteria, \footnotemark\footnotetext{See http://xmm.esac.esa.int/sas/current/documentation/threads/}, and we converted the photon arrival times to the solar system barycenter.

We used {\ttfamily xmmselect} to extract the data products, using the images primarily to create the source and background extraction regions.  The source extraction region is a circle with a $40^{\prime\prime}$ radius centered on the source position.  For EPIC pn, the background region is a rectangle with an area of 1.4 square arcminutes that is located approximately $2^{\prime}$ from the source.  For the MOS detectors, rectangular background regions are also used, but they are farther away from the source because they had to be located on one of the outer MOS CCDs due to the CCD configuration of the small window mode.  The 0.3--12 keV live-time-corrected count rate is 1.3 counts s$^{-1}$ with EPIC pn and 0.8 counts s$^{-1}$ for the two MOS cameras combined.  We checked the observations for pile-up and found that it was not an issue.  

%0.2-15 keV, 6" FWHM resolution (15 arc seconds HEW), 30'x30'
%full frame mode, medium filter MOS - 2.6 s
%pn - 5.6718 ms
%6.5 keV of E/dE~50  (0.1 keV)
%background region: 2' distance from source, 40" radius source, back region area 1.38 arcmin square

\section{Timing Analysis}
\label{sec:timing}

\begin{figure}[b]
\includegraphics[width=0.47\textwidth]{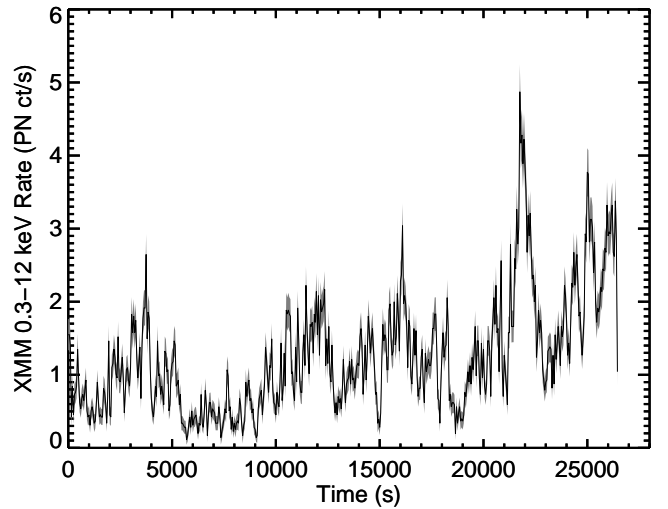}
\caption{Dead-time-corrected light curve from the \textit{XMM-Newton} EPIC pn instrument binned by 50 s.  1$\sigma$ errors are shown in gray.}
\label{fig:xmmlc}
\end{figure}

IGR J18214-1318 exhibits a high level of variability in its X-ray light curve, as can be seen in Figure \ref{fig:xmmlc}.  This strong aperiodic variability is common in HMXBs \citep{belloni90}, and is attributed to variations in the accretion rate resulting from density perturbations due to either disk instabilities or magnetohydrodynamic turbulence (e.g. \citealt{shakura76}; \citealt{hoshino93}; \citealt{revnivtsev09}).  In order to determine whether this HMXB hosts an NS, we searched for periodic pulsations in this noisy light curve through analysis of its power spectrum.  We performed this pulsation search using both \textit{NuSTAR} and \textit{XMM-Newton} EPIC pn data, since the EPIC pn camera has the highest temporal resolution and effective area of the \textit{XMM} instruments.  Although \textit{NuSTAR}'s temporal resolution is better than \textit{XMM}'s, the Earth occultations that \textit{NuSTAR} experiences during its orbit create large gaps in its light curves, which in turn introduce additional noise in the power spectrum at low frequencies.  Furthermore, since pulsars in Sg HMXBs tend to have periods $\gtrsim$1 s (\citealt{corbet86}; \citealt{skinner82}), the 5.7 ms resolution of \textit{XMM-Newton} should be sufficient for detecting possible pulsations in IGR J18214-1318.  We first describe our analysis of the \textit{XMM} EPIC pn power spectrum, and then compare the results to those obtained using \textit{NuSTAR} data. \par
\begin{figure}[t]
\includegraphics[angle=90,width=0.47\textwidth]{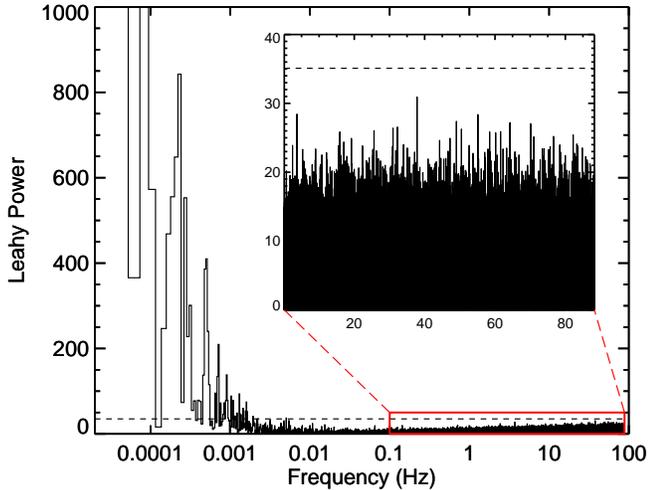}
\caption{Leahy-normalized power spectrum based on the \textit{XMM-Newton} EPIC pn light curve in the 0.3--12 keV band.  The short-dashed black line shows the 90\% confidence detection threshold of 35.1, which is exceeded below $5.06\times10^{-3}$ Hz due to red noise from the source's aperiodic variability.  The inset image shows a zoom-in of the portion of the power spectrum above 0.1 Hz (with a linear frequency scale) so that the highest power peaks in the high-frequency part of the power spectrum can be seen.}
\label{fig:leahy}
\end{figure}
We corrected the arrival time of each event detected by the \textit{XMM-Newton} EPIC pn instrument to the solar system barycenter, and used this corrected event list to make a light curve in the 0.3--12 keV band with the maximum possible time resolution of 5.6718 ms.  We then used the XRONOS tool \texttt{powspec} to produce a Leahy-normalized power spectrum \citep{leahy83} of this light curve, shown in Figure \ref{fig:leahy}.  The power spectrum spans frequencies from 3.7$\times10^{-5}$ Hz (based on the 27 ks duration of the observation) to 88.1 Hz (the Nyquist frequency).  In a Leahy-normalized power spectrum, Poissonian noise results in power being distributed as a $\chi^2$ probability distribution with two degrees of freedom (dof); we used this distribution and the number of trials (which is equal to the number of frequency bins) to calculate the 90\% confidence detection threshold for this power spectrum as 35.1, shown by the dashed lines in Figure \ref{fig:leahy}.  This detection threshold is only exceeded at frequencies $<5.06\times10^{-3}$ Hz, but significant red noise at low frequencies suggests we should be cautious in ascribing this excess power to periodic pulsations.  Thus, in order to account for the red noise present at frequencies $<$0.1 Hz, we analyzed the power spectrum above and below 0.1 Hz separately.  \par
Above 0.1 Hz, the maximum Leahy power ($P_{\mathrm{max}}$) measured is 30.9, which is below the 90\% confidence threshold but can be used to calculate an upper limit on the strength of a periodic signal.  As derived by \citet{vanderklis89}, the 90\% confidence upper limit of the Leahy power ($P_{\mathrm{UL}}$) is given by $P_{\mathrm{UL}}~=~P_{\mathrm{max}}~-~P_{\mathrm{exceed}}$, where $P_{\mathrm{exceed}}$ is the power level exceeded by 90\% of the frequency bins.  In our case, $P_{\mathrm{exceed}}$ is 0.2, which implies that $P_{\mathrm{UL}}$ is 30.7.  This upper limit on the Leahy power can be converted into an upper limit on the source fractional rms variability for a periodic signal using the following formula:
\begin{equation}
\mathrm{rms} = \sqrt{\frac{P_{\mathrm{UL}}-2}{CR}\left(\frac{S+B}{S}\right)^2\Delta\nu}
\end{equation}
where $CR$ is the mean (source plus background) count rate, $S$ represents the net source counts, $B$ represents the estimated background counts in the source region, and $\Delta\nu$ is the width of the frequency bin.  Thus, the 90\% upper limit on the rms noise level for a periodic signal between 0.1 and 88 Hz is $<2.2$\%.  \par

\begin{figure}[t]
\includegraphics[width=0.47\textwidth]{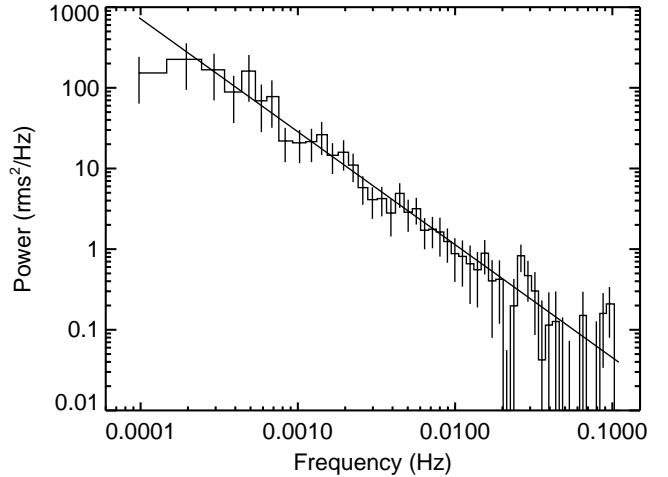}
\caption{The rms-normalized power density spectrum with $1\sigma$ error bars based on the \textit{XMM-Newton} 0.3--12 keV light curve.  The Poisson noise level of 2.0 has been subtracted from the PDS.  The black line shows the best power-law fit to the red-noise-dominated PDS below 0.02 Hz with $\alpha=1.40^{+0.02}_{-0.20}$.}
\label{fig:rms}
\end{figure}

In order to search for periodic signals at low frequencies, we first characterized the red noise below 0.1 Hz.  We produced a 0.3--12 keV light curve with 5 s resolution, and used it to make a rms-normalized power density spectrum (PDS) with a Nyquist frequency of 0.1 Hz and a minimum frequency of 9.8$\times10^{-5}$ Hz, since we averaged together the PDS made from three time intervals of $\sim10$ ks.  The rebinned PDS from which the Poisson noise level of 2.0 has been subtracted is shown in Figure \ref{fig:rms}.  The PDS is dominated by red noise below 0.02 Hz, so we fit the PDS below this frequency with a power-law model ($P = A(\nu/\mathrm{1 Hz})^{-\alpha}$) using Whittle statistics (\citealt{whittle53}; \citealt{whittle57}; \citealt{vaughan10}).  The best-fit parameters are $\alpha = 1.40^{+0.02}_{-0.20}$ and  $A = 0.0018^{+0.006}_{-0.0001}$, where the quoted errors correspond to the 90\% confidence intervals.  The integrated source fractional rms for frequencies between 10$^{-3}$ to 0.1 Hz is $24$\%$\pm1$\%.  The rms PDS slope and integrated fractional rms measured using 3--12 keV \textit{XMM} light curves are consistent with the values measured from the 0.3--12 keV data to better than 1$\sigma$ confidence.  Fitting a power-law model plus a constant to the Leahy-normalized power spectrum from 9.8$\times10^{-5}$ to 88 Hz results also results in a consistent power-law slope.  We also tried fitting a broken power-law model to the rms-normalized and the Leahy-normalized power spectra, but in both cases, the best-fitting break frequency was poorly constrained and exceeded the maximum frequency of the power spectrum.  Thus, we do not find a significant break in the power spectrum between 0.0003 and 88 Hz.  \par
\begin{figure}[t]
\includegraphics[width=0.47\textwidth]{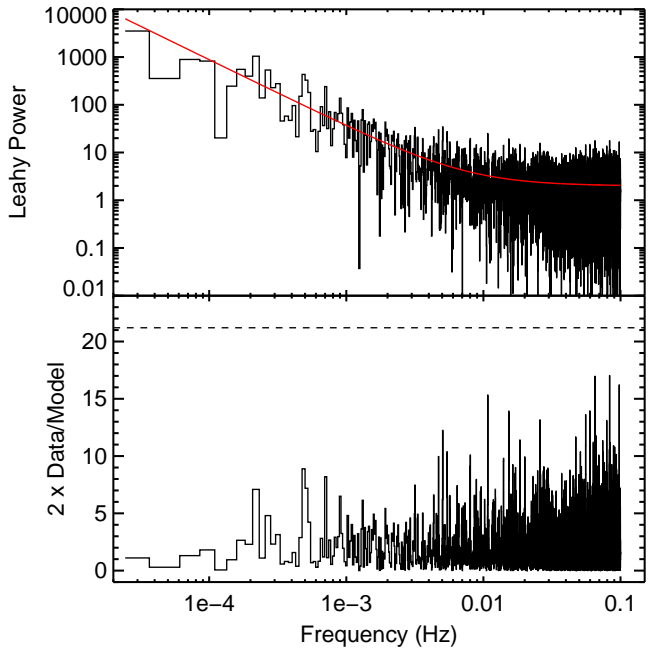}
\caption{The upper panel shows the low-frequency part of the Leahy-normalized power spectrum based on the \textit{XMM-Newton} 0.3--12 keV light curve.  The red solid line shows the appropriately renormalized power-law model derived by fitting the PDS.  The lower panel shows the Leahy power spectrum divided by the model and multiplied by 2.  The dashed line shows the 90\% confidence detection threshold for frequencies below 0.1 Hz, which is equal to 21.2.}
\label{fig:psmodel}
\end{figure}
In order to properly subtract the red noise component from the Leahy-normalized power spectrum in the 0.00003--0.1 Hz range, we follow the procedure described in \citet{vanderklis89}.  The Leahy power spectrum was multiplied by a factor of 2 and then divided by the (appropriately renormalized) best-fit power-law model; we checked that the resulting power spectrum, shown in Figure \ref{fig:psmodel}, follows a $\chi^2$ distribution with 2 dof, a requirement for applying the method described in \citet{vanderklis89} to power spectra with red noise.  Taking into account the number of trials, we calculated the 90\% confidence detection threshold to be 21.2.  As can be seen in Figure \ref{fig:psmodel}, no frequency bin exceeds this power level.  Given that $P_{\mathrm{max}}$ in this low-frequency range is 17.0 and $P_{\mathrm{exceed}}$ is 0.2, $P_{\mathrm{UL}}$ is 16.8, which corresponds to an upper limit on the source fractional rms of 1.7\%.  We verified that varying the red noise slope and normalization within their 90\% confidence intervals does not significantly affect these results. \par
In some binaries, a periodic signal may be difficult to detect because it may be spread out in frequency space due to orbital motion.  However, HMXBs tend to have orbital periods of a few to hundreds of days, and thus our observations of IGR J18214-1318 are short enough that orbital modulations of the pulsation frequency should not be significant.  Thus, given the stringent upper limits on the fractional rms for periodic signals in the 0.00003--88 Hz frequency range, pulsations with periods $\lesssim$1 hr are strongly ruled out.  \par
We performed the same timing analysis with \textit{NuSTAR} data in the 3--12 and 12--30 keV bands.  All \textit{NuSTAR} photon arrival times were converted to barycentric dynamical time (TDB).  After being corrected for thermal drift of the on-board clock, the \textit{NuSTAR} time resolution is $\sim2$ ms rms, and its absolute accuracy is known to be better than 3 ms \citep{mori14}.  Thus, we used the \textit{NuSTAR} data to search for spin periods as short as $\sim$1 ms by binning the light curves by 1/2048 s (488 $\mu$s).  We produced Leahy-normalized power spectra from FPMA and FPMB light curves using both the XRONOS tool \texttt{powspec} and the power spectrum tools developed by M. Bachetti, \footnotemark\footnotetext{Tools can be found at https://bitbucket.org/mbachett/maltpynt and are described in \citet{bachetti15}.} but we did not find any significant peaks in the power spectrum below 1024 Hz.  Above 0.1 Hz, the 90\% confidence upper limit on the source fractional rms is 3.4\% in the 3--12 keV band and 6.4\% in the 12--30 keV band. \par
We produced an rms-normalized, noise-subtracted PDS and fit the red-noise-dominated continuum in the $6\times10^{-5}$ to 0.02 Hz frequency range with a power-law model.  Jointly fitting the PDS produced from FPMA and FPMB data yields $\alpha=1.32^{+0.06}_{-0.10}$ in the 3--12 keV band and $\alpha=1.46^{+0.07}_{-0.12}$ in the 12--30 keV band.  These measured power-law slopes are consistent with the values derived from \textit{XMM-Newton}.  Using the PDS power-law fits to ``normalize'' the red noise continuum in the Leahy power spectrum below 0.1 Hz, we checked that the resulting power spectrum follows a $\chi^2$ distribution with 2 dof, and then calculated that the 90\% confidence upper limit on the source fractional rms is 2.7\% in the 3--12 keV band and 6.0\% in the 12--30 keV band.  The integrated source fractional rms between 10$^{-3}$ to 0.1 Hz is 31\%$\pm$2\% and 23\%$\pm$8\% in the 3--12 and 12--30 keV bands, respectively.  Thus, the \textit{NuSTAR} 3--12 keV integrated fractional rms is higher at 3$\sigma$ confidence than that measured in the \textit{XMM-Newton} 3--12 keV band; this difference may be partly attributed to the \textit{XMM-Newton} and \textit{NuSTAR} observations not being fully coincident in time and the additional artificial noise injected into the \textit{NuSTAR} power spectrum by the light curve gaps due to Earth occultations.  Due to the large errors bars of the integrated fractional rms measured in the \textit{NuSTAR} 12--30 keV band, this value is consistent at 1$\sigma$ confidence with both the integrated rms measured by \textit{XMM-Newton} and the \textit{NuSTAR} 3--12 keV band.  \par
Overall, the results from our \textit{XMM-Newton} and \textit{NuSTAR} timing analyses are in agreement, ruling out the presence of pulsations with periods shorter than about an hour.  The integrated source fractional rms values measured between $10^{-3}$ and 0.1 Hz by both telescopes are within the typical range of 10-30\% seen in HMXBs \citep{belloni90}.  For most accreting X-ray pulsars, the red noise power-law index is $\alpha=1.4-2.0$ at frequencies higher than the pulsation frequency and $\alpha = 0-1.0$ at lower frequencies \citep{hoshino93}.  Thus, the red noise power-law slope of IGR J18214-1318 is similar to the slopes observed above the pulsation frequency in X-ray pulsars.  This fact, combined with the lack of a frequency break in the red noise continuum, suggests, but does not prove, that the pulsation frequency in IGR J18214-1318 may be lower than the range probed by our data.

\section{Spectral Analysis}
\label{sec:spectral}

\begin{figure*}
\makebox[\textwidth]{ %
	\centering
	\subfigure{\includegraphics[width=0.48\textwidth]{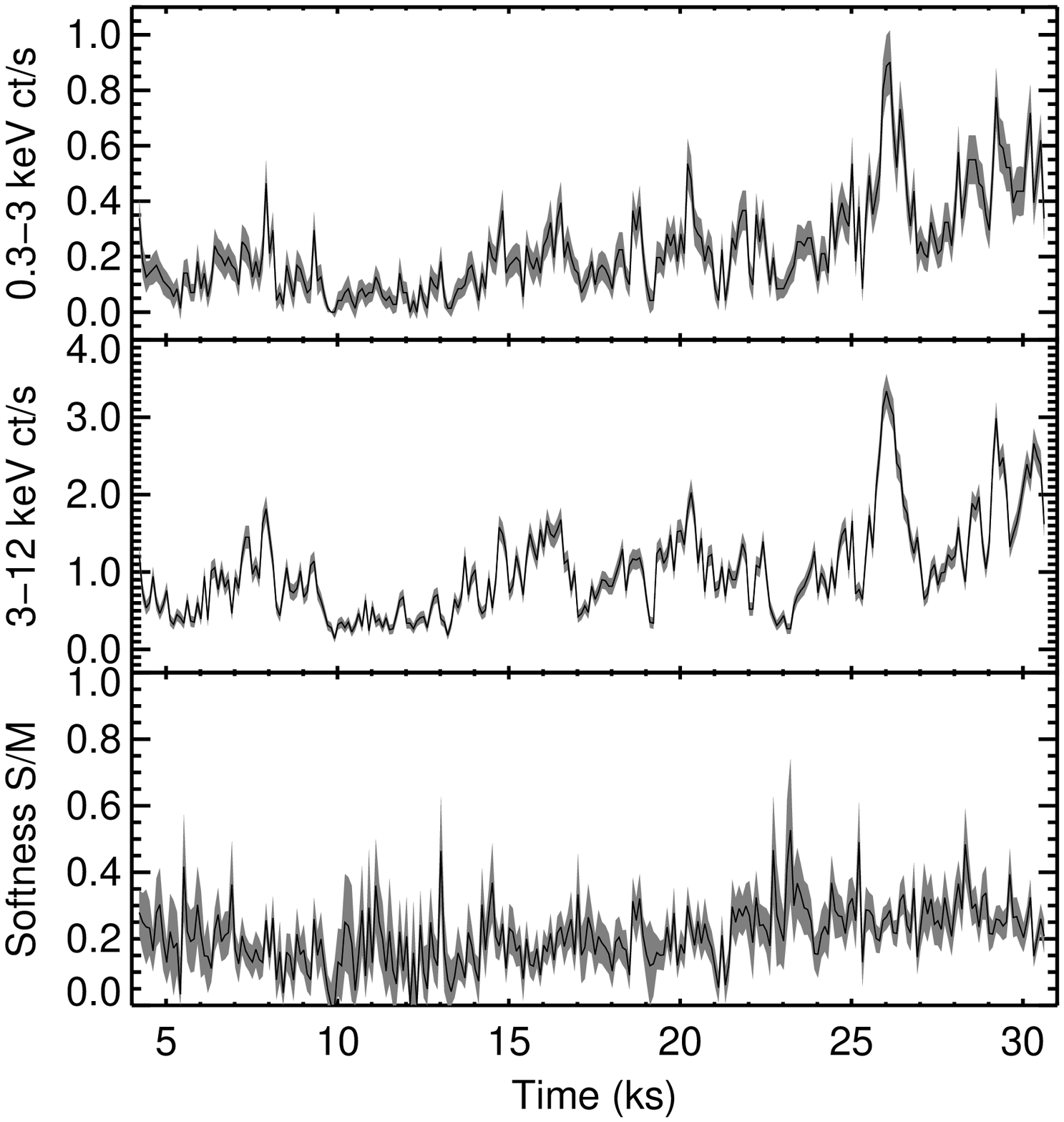}}
	\hspace{0.1in}\subfigure{\includegraphics[width=0.49\textwidth]{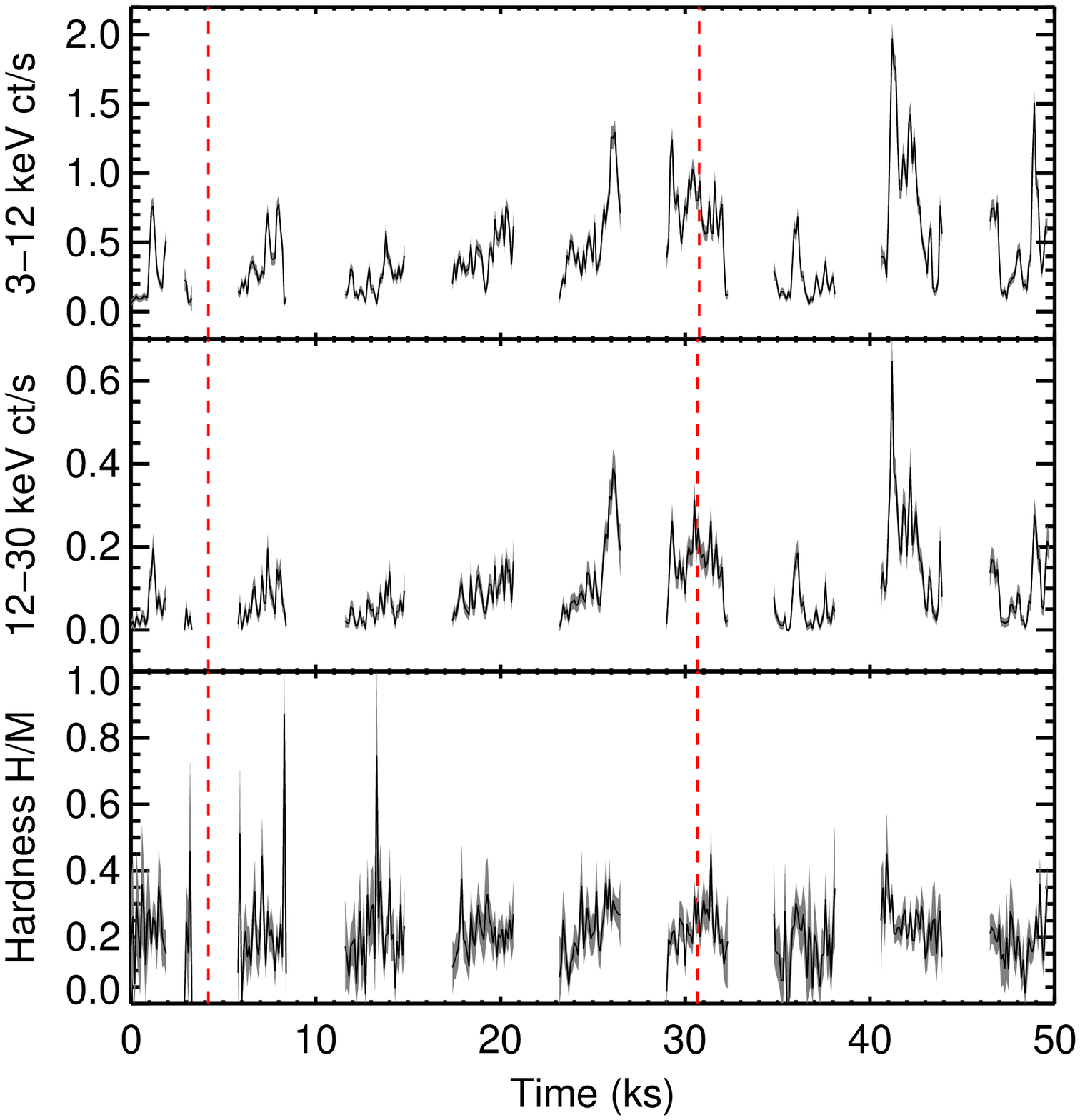}}}
\caption{\textit{Left:} \textit{XMM-Newton} EPIC pn light curves in the 0.3--3 and 3--12 keV bands binned by 100 s.  The softness ratio in the lower panel is defined as $S/M$, where $M$ is the count rate in the medium 3--12 keV band and $S$ is the count rate in the soft 0.3--3 keV band.  Time on the x-axis is measured from the beginning of the \textit{NuSTAR} observations.  \textit{Right:} \textit{NuSTAR} lightcurves, FPMA and FPMB combined, binned by 100 seconds.  The hardness ratio is calculated as $H/M$, where $H$ is the count rate in the hard 12--30 keV band.  The gaps in the \textit{NuSTAR} light curves are due to Earth occultations.  The red dashed lines indicate the beginning and end of the \textit{XMM-Newton} observations.}
\label{fig:lcratio}
\end{figure*}

Since IGR J18214-1318 exhibits strong variability, when performing spectral analysis it is important to consider whether the source spectrum varies with source brightness.  First, we checked for substantial spectral variations correlated with source flux by producing light curves in soft, medium, and hard energy bands (0.3--3, 3--12, and 12--30~keV, respectively), and using them to calculate hardness/softness ratios as a function of time, which are shown in Figure \ref{fig:lcratio}.  With the \textit{NuSTAR} data, we calculate the hardness ratio as the count rate in the hard energy band divided by the rate in the medium band.  With the \textit{XMM-Newton} data, we calculate the softness ratio as the count rate in the soft energy band divided by the rate in the medium energy band.  The fact that the \textit{XMM-Newton} softness ratio and the \textit{NuSTAR} hardness ratio have the same energy band in the denominator helps to visualize how the soft and hard energy ends of the X-ray spectrum vary relative to one another. The average ratios in both the \textit{XMM-Newton} and \textit{NuSTAR} bands do not vary significantly on timescales of $3-4$~ks, remaining consistent within 1$\sigma$ confidence intervals.  However, there are indications of ratio variability on shorter timescales, as several \textit{XMM-Newton} and \textit{NuSTAR} ratios (5 of 265 and 10 of 287, respectively) measured in 100s intervals differ from the mean ratio by $>2.7\sigma$, even though only one ratio measurement per instrument is statistically expected to differ by this amount.
%\textbf{As shown in Figure \ref{fig:lcratio}, the ratios in both the \textit{XMM-Newton} and \textit{NuSTAR} bands remain fairly constant throughout the observations.  However, the average statistical uncertainties of $\pm$0.1 are comparable to the range of values spanned by the ratio measurements, with 80\% of both \textit{XMM-Newton} and \textit{NuSTAR} ratios falling between 0.1 and 0.3; such relatively large uncertainties could wash out variations of the hardness/softness ratios with count rate. } \par

\begin{figure*}
\makebox[\textwidth]{ %
	\centering
	\subfigure{
		\includegraphics[width=0.47\textwidth]{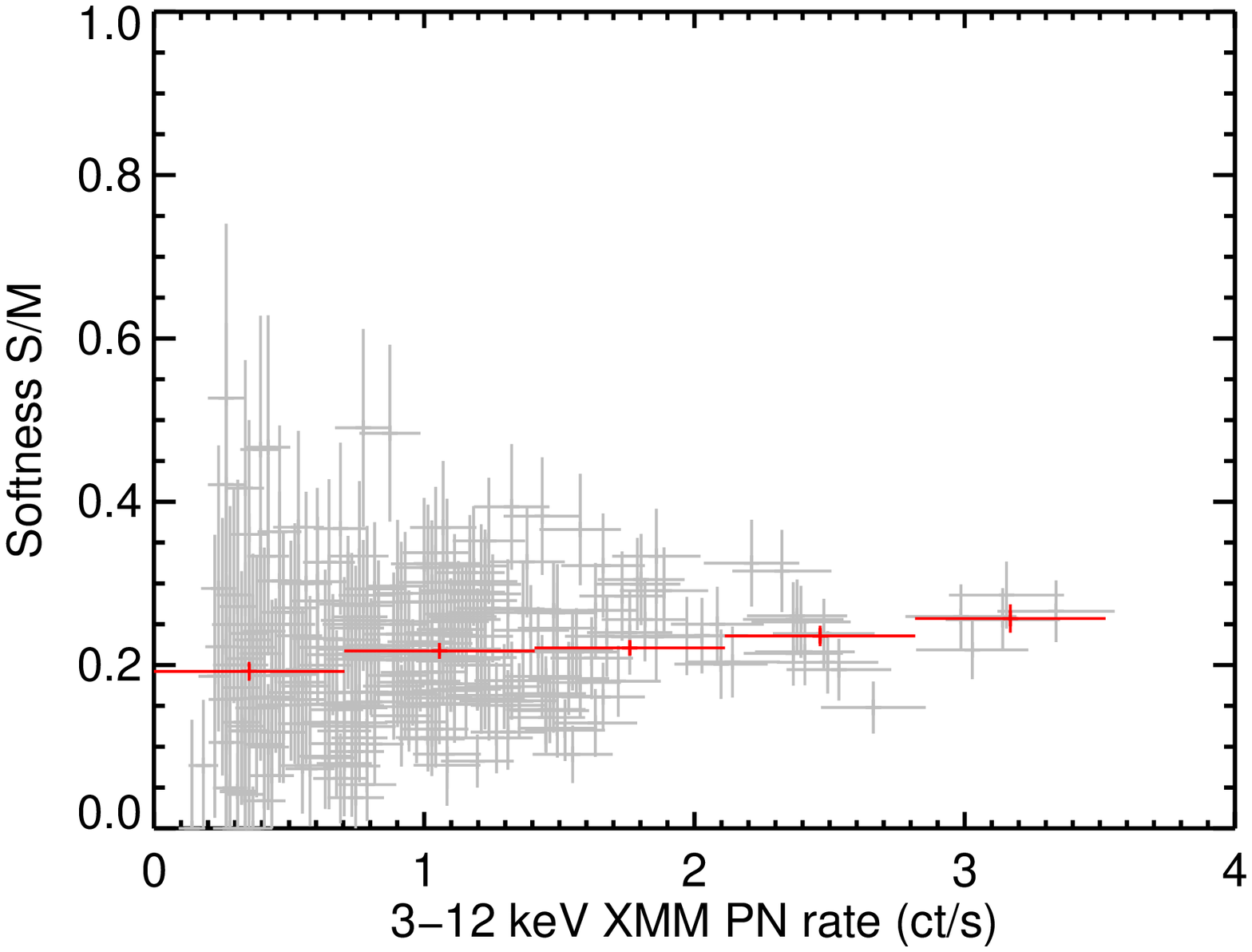}}
	\hspace{0.1in}\subfigure{
		\includegraphics[width=0.47\textwidth]{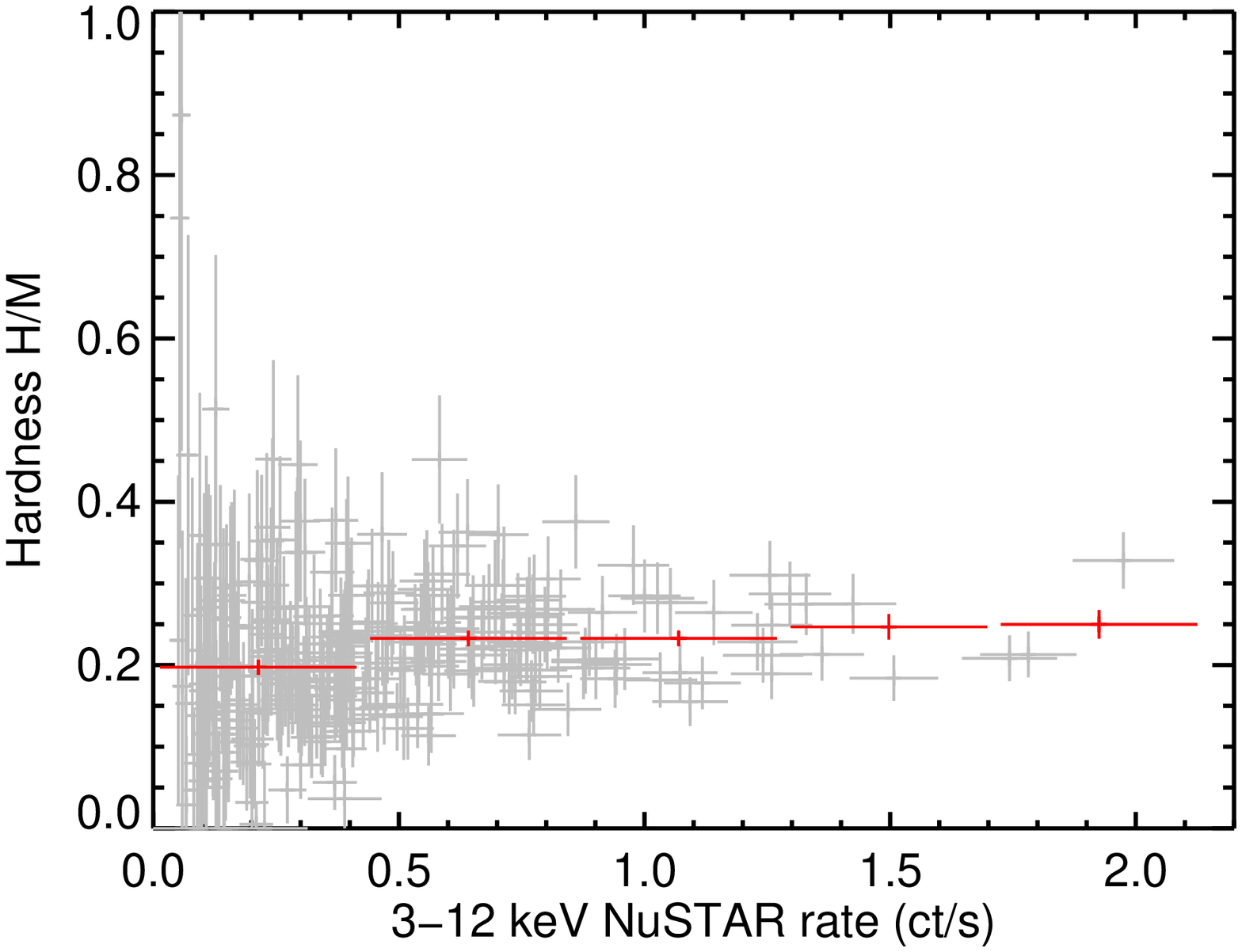}}}
\caption{The left panel shows the \textit{XMM-Newton} softness ratio versus 3--12 keV count rate, while the right panel shows the same for \textit{NuSTAR} data.  Gray points show data from 100s light-curve bins.  Red points show the average hardness/softness ratio in five count rate bins.}
\label{fig:ratio}
\end{figure*}

In order to establish whether there is a significant correlation between the source brightness and X-ray ratios, we grouped together the individual measurements from 100 s light-curve fragments with similar count rates, creating a total of five count rate bins and evaluating the mean X-ray hardness/softness ratio in each bin.  The standard deviation of the distribution of ratio values within each count rate bin differs by $<15$\% from the average 1$\sigma$ uncertainty of the ratio measurements in that bin, indicating that the apparent scatter in ratios within each count rate bin is primarily statistical, not intrinsic, in nature; thus, averaging the ratio values within a given count rate bin reduces the statistical error, which for each individual ratio measurement is $\pm0.1$ on average.  We calculate the error on the mean as the quadrature sum of the individual measurement uncertainties divided by the total number of measurements averaged together. \par
As shown in Figure \ref{fig:ratio}, in both the \textit{XMM-Newton} and \textit{NuSTAR} bands, the mean X-ray ratio increases slightly with increasing count rate, as is observed in some other Sg HMXBs (e.g., IGR~J16207$-$5129 and 4U~2206+54; \citealt{tomsick09}; \citealt{wang13}).  Fitting a constant value to the mean ratios in Figure \ref{fig:ratio} using $\chi^2$ minimization, we find that the null hypothesis that the ratios do not vary with count rate is rejected at 98.4\% confidence in the \textit{XMM-Newton} band and at 99.9\% confidence in the \textit{NuSTAR} band.  While the variations of X-ray ratios with count rate are statistically significant, the spectral variations of IGR~J18214-1318 are likely to be small given that the mean X-ray ratios increase by only 25\% as the count rate increases by almost an order of magnitude; therefore, it is unlikely that significant biases would be introduced by fitting the average source spectrum derived from all available data.  Nonetheless, verifying whether the flux-dependent spectral variations are indeed small and determining the cause of variations, if they exist, can be valuable for understanding the origin of variability in IGR~J18214-1318.  Therefore, we made one spectrum, which we refer to as the flux-averaged spectrum, combining all \textit{XMM-Newton} and \textit{NuSTAR} data to achieve maximum sensitivity to detect potentially weak features such as emission lines or cyclotron absorption lines, and we investigated the origin and significance of spectral variations by making a low flux and a high flux spectrum based on simultaneous intervals of \textit{XMM-Newton} and \textit{NuSTAR} observations. 
%Thus, as the source flux increases, the relative contributions of the 0.3-3, 3-12, and 12-30 keV bands to the total absorbed flux become more comparable, resulting in an overall flattening of the spectral shape,}
%The ratio trends imply} that as the source flux increases, the source spectrum between 0.3 to 30~keV becomes flatter

\subsection{The flux-averaged spectrum}
\label{sec:avgspectrum}
First, we describe the spectral analysis of the flux-averaged source spectrum.  We extracted spectra, ARFs, and RMFs from the \textit{XMM-Newton} EPIC pn, MOS1/2, and \textit{NuSTAR} FPMA/B instruments as described in \S\ref{sec:obs}.  The spectra were rebinned with the requirement that the signal significance in each bin be $\geq10$, except for the highest energy bin which was required to have a significance $\geq3$.  We used the XSPEC version 12.8.2 software to jointly fit the five spectra, allowing for different calibration constants for each instrument.  The cross-calibration constants for the MOS1 and MOS2 instruments were consistent in all the fits, differing by less than 1$\sigma$ from each other, so we linked the MOS1/2 constants together, removing one free parameter from the models.  \par

\begin{figure}[t]
\includegraphics[angle=270,width=0.47\textwidth]{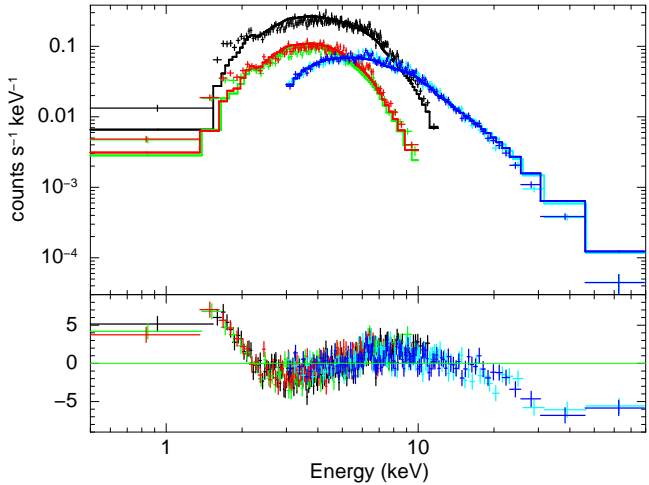}
\caption{Spectrum of IGR J18214-1318, with \textit{XMM-Newton} EPIC pn data shown in black, MOS1 in green, MOS2 in red, \textit{NuSTAR} FPMA in light blue, and FPMB in dark blue.  Data points are shown with 1$\sigma$ error bars.  Fit shown is for a simple absorbed power-law model.  The lower panel shows the residuals to the fit, where $\chi$ = (data-model)/error. The residuals in the lower panel show there is an excess below 2 keV and a flux deficit above 10 keV. }
\label{fig:simplepl}
\end{figure}

\begin{figure*}
\makebox[\textwidth]{ %
	\centering
	\subfigure{\includegraphics[angle=270,width=0.47\textwidth]{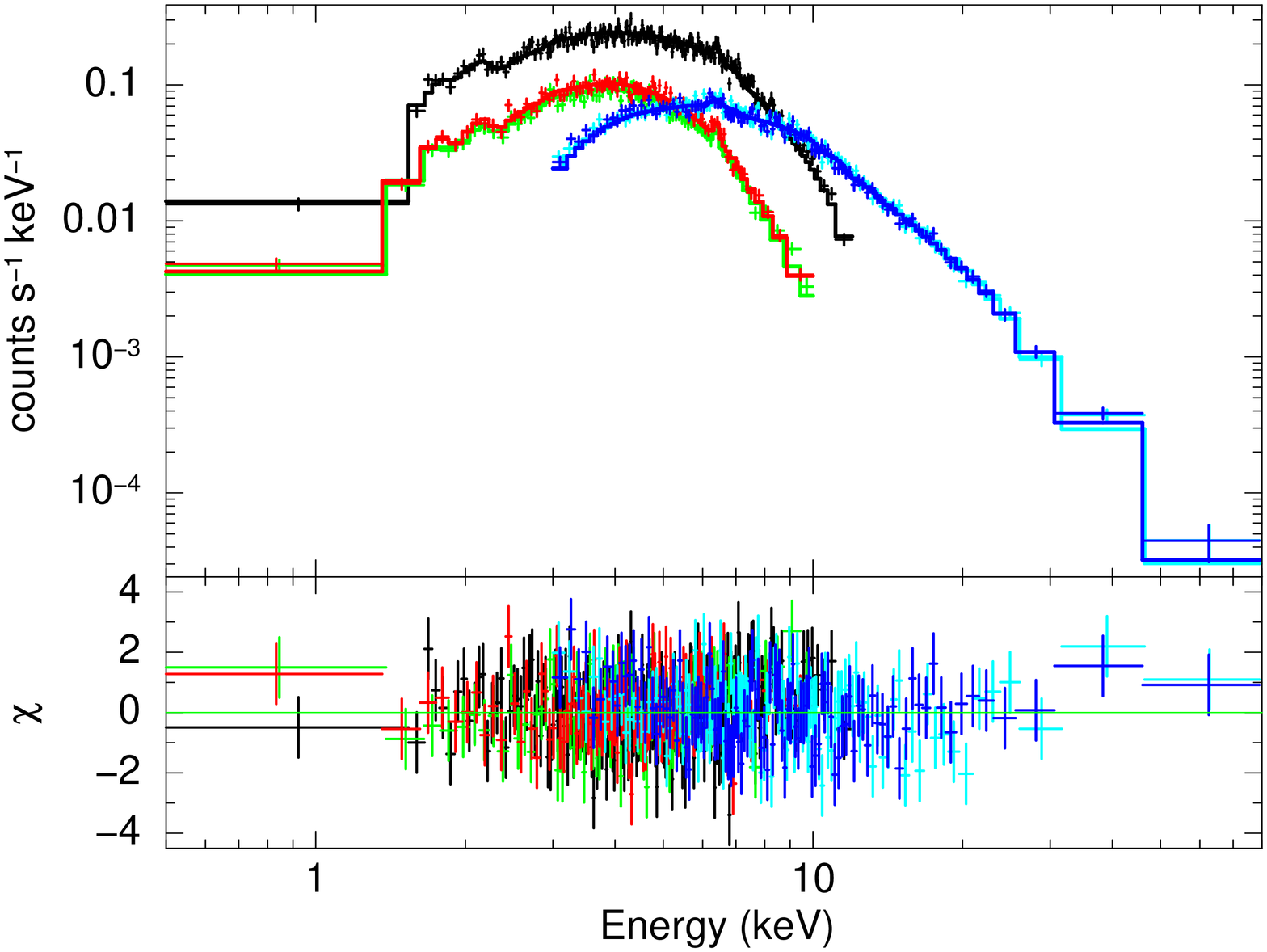}}
	\hspace{0.1in}\subfigure{\includegraphics[angle=270,width=0.46\textwidth]{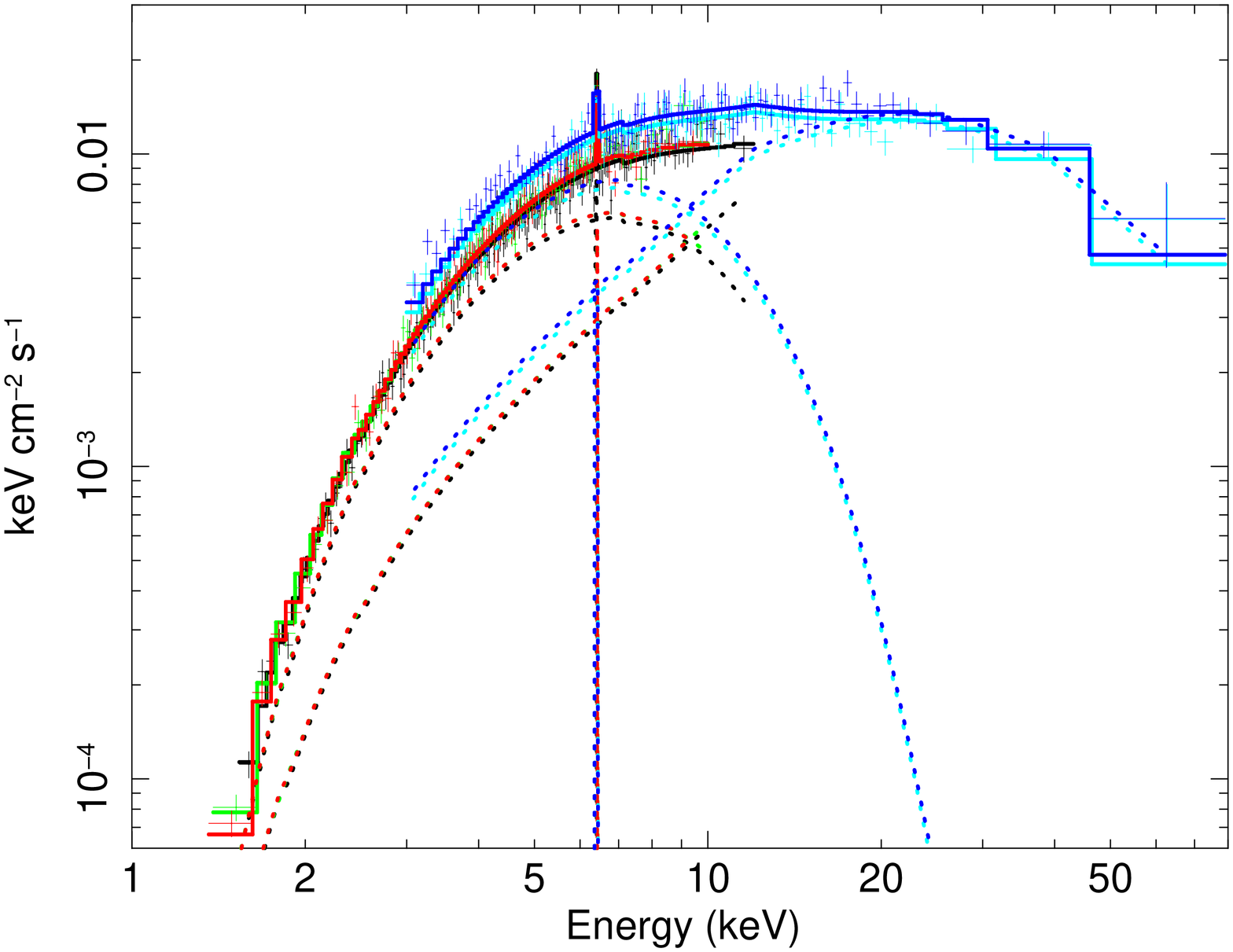}}}
\caption{\textit{Left}: The model fit shown is for an absorbed power law with a high-energy cutoff, a partial-covering absorber, and a Gaussian Fe line.  \textit{Right:} The spectral energy density is shown with a model including blackbody and cutoff power-law components subject to the same absorption.  A Gaussian line to account for Fe K$\alpha$ emission is also included. The different model components are shown with dashed lines.  The colors of the data points shown are as described in Figure \ref{fig:simplepl}.}
\label{fig:goodfit}
\end{figure*}

\begin{table}
\centering
\footnotesize
\caption{Best-fit Spectral Parameters}
\begin{threeparttable}
\begin{tabular}{cccccccc} \hline \hline
\T& Model 1 & Model 2 \\
&\texttt{tbabs*(bbody+}&\texttt{tbabs*pcfabs*}\\
&\texttt{powerlaw*highecut}&(\texttt{powerlaw*highecut}\\
\B &\texttt{+gaussian)}&\texttt{+gaussian)}\\
\hline
\T\B $N_{\mathrm{H}}$ ($10^{22}$ cm$^{-2}$) & 4.2$^{+0.3}_{-0.2}$ & 4.3$\pm$0.6 \\
\B $kT_{\mathrm{BB}}$ / $N_{\mathrm{H,partial}}$ & 1.74$^{+0.04}_{-0.05}$ keV & 9.8$^{+1.5}_{-1.1} \times 10^{22}$ cm$^{-2}$ \\
\B BB norm. / Cov. frac. & 1.3$\pm0.1 \times 10^{-11}$ & 0.77$^{+0.05}_{-0.06}$ \\
\B $\Gamma$ & 0.4$^{+0.3}_{-0.4}$ & 1.48$^{+0.08}_{-0.07}$\\
\B PL norm.  & 4.9$\pm2.0 \times 10^{-12}$ & 2.6$\pm0.2 \times 10^{-11}$\\
\B $E_{\mathrm{cut}}$ (keV) & 12.0$^{+1.0}_{-1.3}$ & 7.4$^{+0.6}_{-0.5}$\\
\B $E_{\mathrm{fold}}$ (keV) & 14.0$^{+3.2}_{-1.5}$ & 23.0$^{+3.3}_{-2.4}$\\
\B $E_{\mathrm{line}}$ (keV) & 6.40$^{+0.03}_{-0.02}$ & 6.40$\pm$0.02\\
\B $\sigma_{\mathrm{line}}$ (eV) & $<85$ & $<102$\\
\B EW$_{\mathrm{line}}$ (eV) & 53$^{+16}_{-21}$ & 57$^{+16}_{-14}$\\
\B $C_{\mathrm{MOS1,2}}$ & 1.04$\pm$0.02 & 1.04$\pm$0.02\\
\B $C_{\mathrm{FPMA}}$ & 1.25$\pm$0.02 & 1.24$\pm$0.02\\
\B $C_{\mathrm{FPMB}}$ & 1.32$\pm$0.03 & 1.31$\pm$0.03\\
\B $\chi^2_{\nu}$/dof & 1.10/578 & 1.10/578\\
\hline
\end{tabular}
\begin{tablenotes}[flushleft]
\item Notes: Errors provided are 90\% confidence.  Cross-normalizations between instruments are calculated relative to the \textit{XMM-Newton} EPIC pn instrument.  Abbreviations: BB--blackbody, PL--power law.  The BB normalization is the unabsorbed 0.3--10~keV flux (in units of erg~cm$^{-2}$~s$^{-1}$) of the BB component.  The PL normalization is the 0.3--10~keV flux (in units of erg~cm$^{-2}$~s$^{-1}$) of the PL component (in the case of the \texttt{pcfabs} model, it is the flux of the PL component that is subject to local absorption by the partial-covering absorber). 
%The BB normalization is the source luminosity in units of $10^{39}$ erg s$^{-1}$ assuming a distance of 10 kpc.  The PL normalization is the photon flux at 1 keV.  
\end{tablenotes}
\end{threeparttable}
\label{tab:spectra}
\end{table}

\begin{figure}[b]
\includegraphics[angle=270,width=0.47\textwidth]{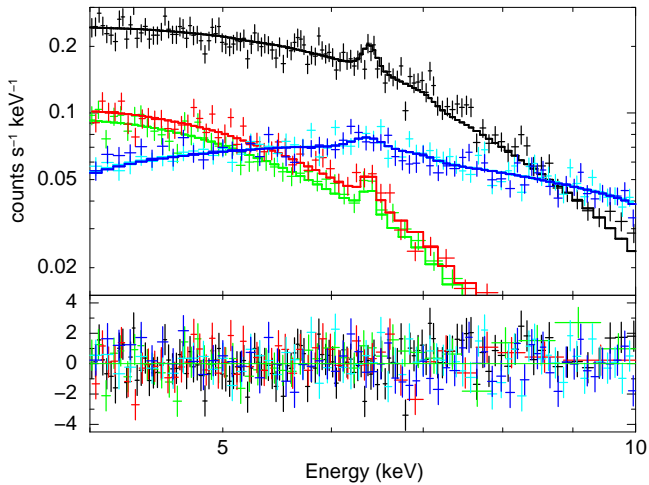}
\caption{Zoom-in of the spectrum from 4 to 10 keV band clearly shows the presence of an Fe line around 6.4 keV.  The colors of the data points shown are as described in Figure \ref{fig:simplepl}.}
\label{fig:feline}
\end{figure}

We first fit the data using an absorbed power-law model (\texttt{tbabs*powerlaw}), adopting the abundances from \citet{wilms00} and photoionization cross-sections from \citet{verner96}.  This simple model, which was sufficient for describing previously available soft X-ray data with lower photon statistics (\citealt{tomsick08}; \citealt{rodriguez09}), yields a poor fit ($\chi^2_{\nu}$ = 3.3 for 585 dof).  As can be seen in the residuals in Figure \ref{fig:simplepl}, this simple power-law fit underestimates the flux below 2 keV and overestimates it above 20 keV.  Accounting for the flux above 20 keV requires introducing an exponential cutoff to the power-law spectrum, while the soft excess can be accounted for either by adding a blackbody component (Model 1) or a partial-covering absorber (Model 2), which provide equally good fits.  Adding only one of these components (\texttt{highecut}, \texttt{bbody}, or \texttt{pcfabs}) to the absorbed power-law model is insufficient, leaving large residuals either below 2 keV or above 20 keV. \par
The spectral fits and residuals resulting from our best-fitting models are shown in Figure \ref{fig:goodfit}.   These models also include a Gaussian line to fit the Fe K$\alpha$ line emission at $6.40\pm0.02$ keV, which is clearly visible in Figure \ref{fig:feline}.  The energy of this line indicates it must originate in cool, low-ionization material located in the supergiant wind \citep{torrejon10}.  The spectral parameters of the best-fitting models are listed in Table \ref{tab:spectra}.  As can be seen, the reduced $\chi^2$ values of the spectral fits are good enough that no additional components are required, and no prominent features remain in the residuals. \par  
However, in order to statistically test for the presence of cyclotron lines, we added a cyclotron absorption component (\texttt{cyclabs}) to our models and performed new spectral fits.  Since the cyclotron line width was very poorly constrained when left as a free parameter, we set its upper limit to 10 keV, since cyclotron line widths of accreting X-ray pulsars typically fall in the 1--10 keV range \citep{coburn02}.  We also set the optical depth of the second harmonic to zero since it could not be constrained.  The spectral parameters of all the other Model 1 and 2 components were allowed to vary in order to find the best-fitting model that includes \texttt{cyclabs} as a multiplicative component.  The best-fitting cyclotron line parameters derived by adding \texttt{cyclabs} to Model 1 are an optical depth $\tau_{\mathrm{cyc}} = 0.25^{+0.21}_{-0.18}$ and line energy $E_{\mathrm{cyc}} = 27^{+4}_{-6}$ keV; the inclusion of \texttt{cyclabs} only reduced the chi-squared value of the fit by 5 and left $\chi^2_{\nu}$ unchanged.  The cyclotron parameters derived by adding \texttt{cyclabs} to Model 2 are $\tau_{\mathrm{cyc}} = 0.16\pm0.06$ and $E_{\mathrm{cyc}} = 11^{+4}_{-7}$ keV, which differs from the cyclotron energy found for Model 1; adding the \texttt{cyclabs} component to Model 2 improved the chi-squared value by 19.2 and reduced $\chi^2_{\nu}$ to 1.07 from 1.10, a marginal improvement on the quality of the fit.  \par
In order to determine the significance of this improvement to the chi-squared value for Model 2, we  generated 1000 simulated datasets, including both the \textit{NuSTAR} and \textit{XMM} data and followed the procedure applied in \citet{bellm14}, \citet{bhalerao15}, and \citet{bodaghee16}.  Each simulated dataset was fit by the null model (Model 2 without \texttt{cyclabs}) and the test model with a \texttt{cyclabs} feature, and the difference in chi-squared values ($\Delta\chi^2$) between the two model fits was calculated.  The maximum value of $\Delta\chi^2$ from these simulations was 19.3, slightly higher than the observed value.  Based on the distribution of $\Delta\chi^2$ from our simulations, we estimate there is roughly a 0.001\% chance of measuring the observed value of $\Delta\chi^2=19.2$ by chance, and that therefore the significance of the cyclotron line in IGR J18214-1318 is about $3.3\sigma$.  
%In the case of both Model 1 and Model 2, fixing the cyclotron line width to other values within the typical range of 1--10 keV \citep{coburn02} does not significantly alter the best-fitting $E_{\mathrm{cyc}}$ or $\tau_{\mathrm{cyc}}$ or result in a lower $\chi^2_{\nu}$.  
Given the fact that this detection is marginal and dependent on adopting Model 2 rather than Model 1 for the soft excess, it does not constitute substantive evidence for the presence of a cyclotron absorption feature. \par 
Nonetheless, the absence of such features does not disprove the possibility that IGR J18214-1318 harbors an NS.  The fact that the e-folding energy of the exponential cutoff is $<25$ keV, regardless of which of the two best models is adopted, strongly suggests that IGR J18214-1318 is an NS HMXB.  Furthermore, the photon index below the cutoff is harder when adopting the blackbody rather than the partial-covering model, but in both cases is within the range observed in NS HMXBs, which tend to exhibit harder photon indices than BH HMXBs \citep{coburn02}.  \par
For Model 1, the blackbody component accounting for the soft excess has a temperature of 1.7 keV, which is higher than the $kT\approx0.1$ keV thermal component exhibited by BH HMXBs in the hard state (\citealt{disalvo01}; \citealt{mcclintock06}; \citealt{makishima08}); BH HMXBs can exhibit blackbody temperatures as high as 2 keV in the soft state, but the power-law component of IGR J18214-1318 is much stronger than that of a BH in the soft state \citep{mcclintock06}.  Assuming that IGR J18214-1318 lies at a distance of  9--10~kpc, as favored by the properties of its near-IR counterpart \citep{butler09}, the radius of the blackbody-emitting region is 0.3~km, which is consistent with the size of NS hot spots.  However, while the blackbody interpretation thus provides some additional evidence in favor of the NS hypothesis, the soft excess seen in the spectrum can be equally well-fit by a partial-covering model.  Using this model, we measure that the whole system lies behind a column density of $4\times10^{22}$ cm$^{-2}$, which is just in excess of the Galactic interstellar column density integrated along the line of sight, and that about 77\% of the X-ray emission is obscured by an additional column density of $\sim10^{23}$ cm$^{-2}$.  This partial-covering absorber can be attributed to dense clumps in the supergiant wind, and thus does not provide any additional insight into the compact object in this HMXB.  \par

\begin{figure*}
\makebox[\textwidth]{ %
	\centering
	\subfigure{\includegraphics[angle=270,width=0.48\textwidth]{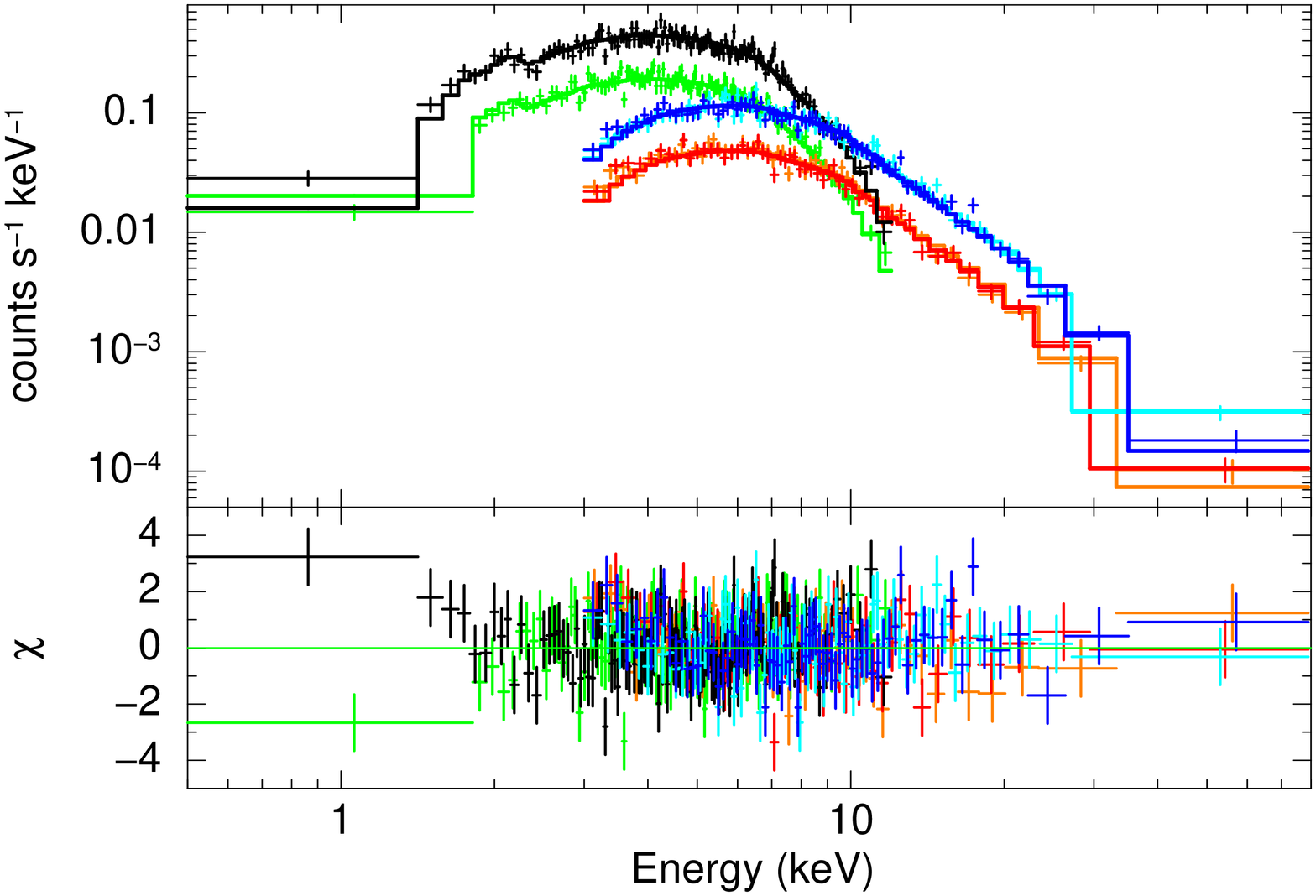}}
	\hspace{0.1in}\subfigure{\includegraphics[angle=270,width=0.48\textwidth]{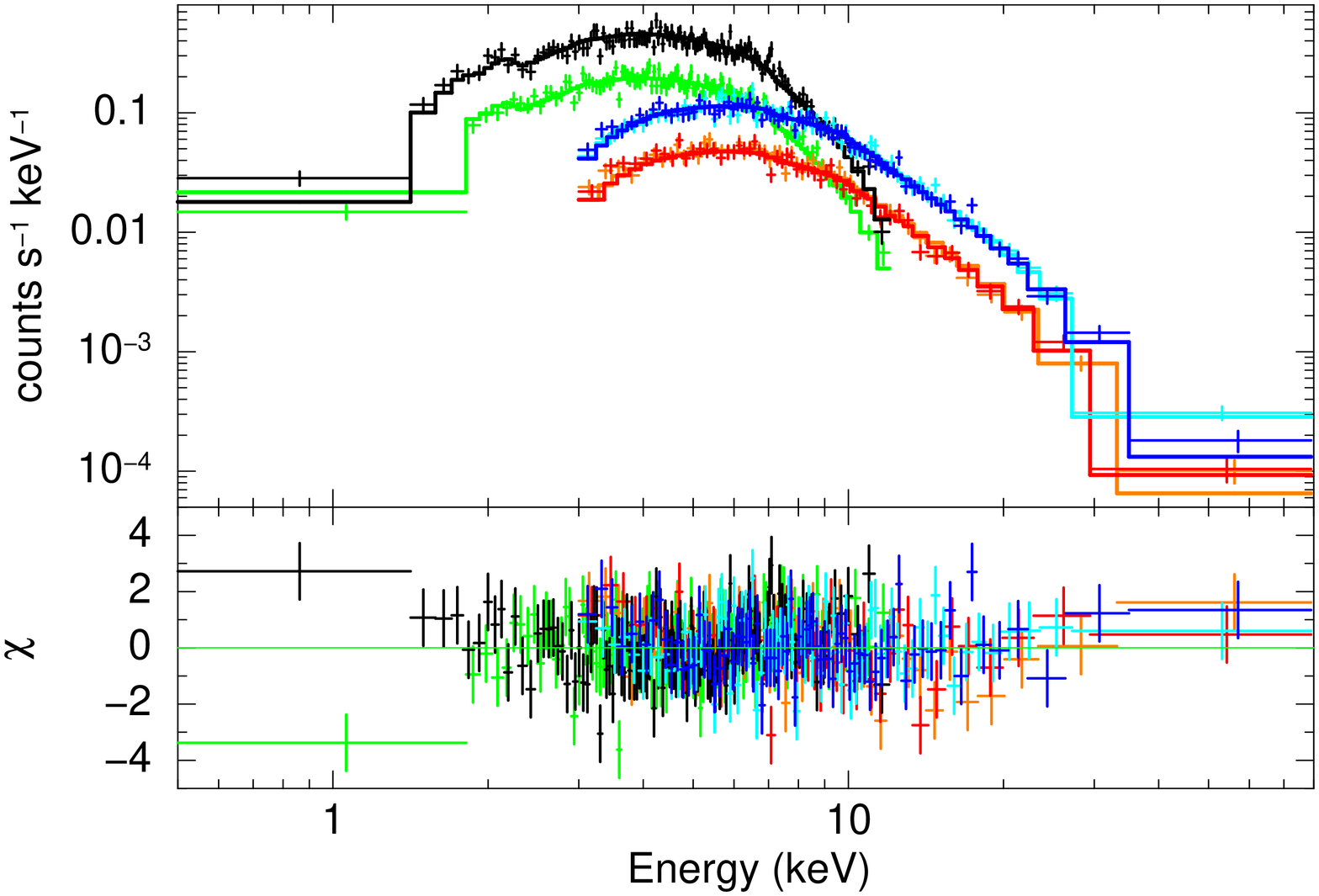}}}
\caption{Low (high) flux spectra of IGR J18214-1318, with \textit{XMM-Newton} EPIC pn data shown in green (black), \textit{NuSTAR} FPMA in orange (light blue), and FPMB in red (dark blue).  Data points are shown with 1$\sigma$ error bars.  Fit shown is for an absorbed power-law model with a high-energy cutoff plus a blackbody component (left) or a partial-covering absorber (right).  In the best-fit models shown, the normalization constants for the individual spectra (for different instruments and flux states) are left as free parameters, but all other model parameters are tied together while fitting the low and high flux state spectra. The fit residuals for the high and low flux state spectra only differ significantly at energies below 2~keV. }
\label{fig:specstates}
\end{figure*}

The mean 3--12 keV flux is 1.70$^{+0.02}_{-0.05}\times10^{-11}$ erg~cm$^{-2}$~s$^{-1}$ in the \textit{XMM-Newton} EPIC pn observations and 2.18$^{+0.03}_{-0.10}\times10^{-11}$ erg~cm$^{-2}$~s$^{-1}$ in the \textit{NuSTAR} observations (FPMA and FPMB averaged).  At a distance of 9--10 kpc, these fluxes correspond to unabsorbed luminosities of $\approx1-2\times10^{35}$~erg~s$^{-1}$.  Through simultaneous \textit{NuSTAR} and \textit{XMM-Newton} observations of PKS~2155-304 and 3C~273, \citet{madsen15} found that the flux cross-calibrations of the \textit{XMM-Newton} and \textit{NuSTAR} instruments are accurate to better than 7\%; in our spectral fits, the \textit{NuSTAR} fluxes (see Table \ref{tab:spectra}) are higher than the \textit{XMM-Newton} fluxes by about 30\%, primarily because the \textit{NuSTAR} observations cover a longer duration of time and the source undergoes some large flares after the \textit{XMM-Newton} observations end, as shown in Figure \ref{fig:lcratio}.  The average X-ray flux during these observations is a factor of 3 lower than the average flux during the 2008 and 2009 observations (\citealt{tomsick08}; \citealt{rodriguez09}), and the light curves in Figure \ref{fig:lcratio} show that on hour-long timescales, the flux can vary by more than a factor of 20. 

\subsection{Spectral variations with brightness}
In order to verify whether the flux-dependent spectral variations are small in IGR~J18214-1318, as suggested by the stability of the hardness ratios with time, and to investigate the origin of the spectral variations if they exist, we made low flux state and high flux state spectra.  The low (high) flux \textit{XMM-Newton} EPIC pn and \textit{NuSTAR} FPMA/B spectra consist of counts extracted from 100s time intervals of simultaneous \textit{XMM-Newton} and \textit{NuSTAR} observations during which the \textit{NuSTAR}, FPMA and FPMB combined, 3--12~keV count rate is less (greater) than 0.5 counts s$^{-1}$.  This count rate threshold was chosen taking into account the point at which the hardness ratio changes most substantially in Figure \ref{fig:ratio} and trying to ensure that the low and high state spectra would have roughly the same number of total source counts.  The spectra were rebinned so that the signal significance in each energy bin was $\geq7$, except for the highest energy bin which was only required to have a significance $\geq3$. \par
First, we fit the low and high state spectra independently with a simple absorbed power-law model.  Both the low and high state spectra exhibit data residuals to the model fit that are very similar to those of the flux-averaged spectrum shown in Figure \ref{fig:simplepl}, with an excess below 2~keV and a deficit above 10~keV.  In both cases, in order to remove the structure in these residuals and produce a good $\chi_{\nu}^2$ value, it is necessary to both add a high-energy cutoff to the power law and to introduce either a blackbody or a partial-covering absorber to model the soft excess.  Thus, both the soft excess and high-energy cutoff appear to be persistent features of the spectrum of IGR~J18214-1318.  The addition of a Gaussian component does not significantly improve the $\chi_{\nu}^2$ value of the fits to either the low or high state spectra, as the sensitivity of the low and high state spectra is lower than that of the flux-averaged spectrum and insufficient to detect the relatively weak Fe line at 6.4~keV. \par

\begin{figure*}
\makebox[\textwidth]{ %
	\centering
	\subfigure{\includegraphics[width=0.48\textwidth]{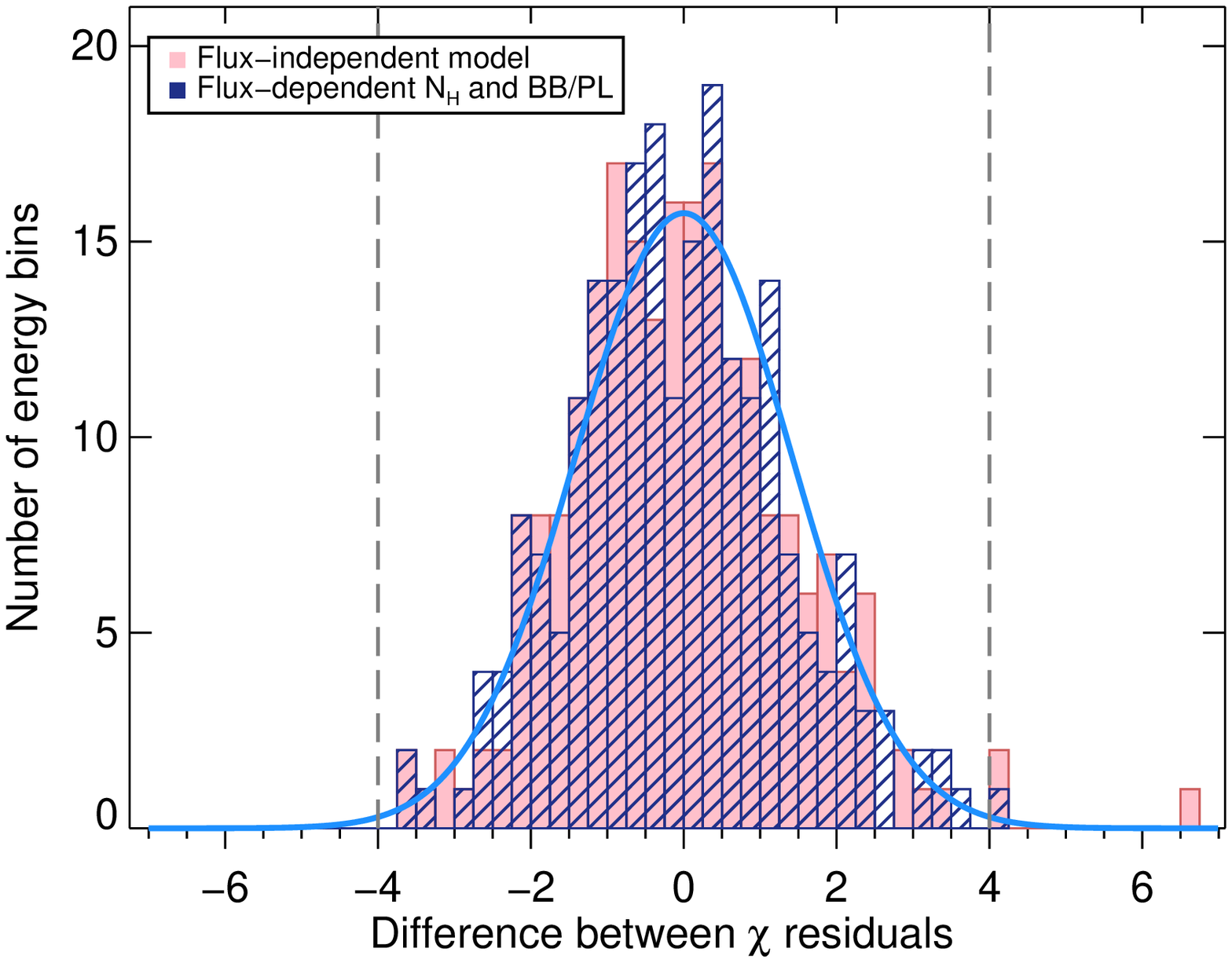}}
	\hspace{0.1in}\subfigure{\includegraphics[width=0.48\textwidth]{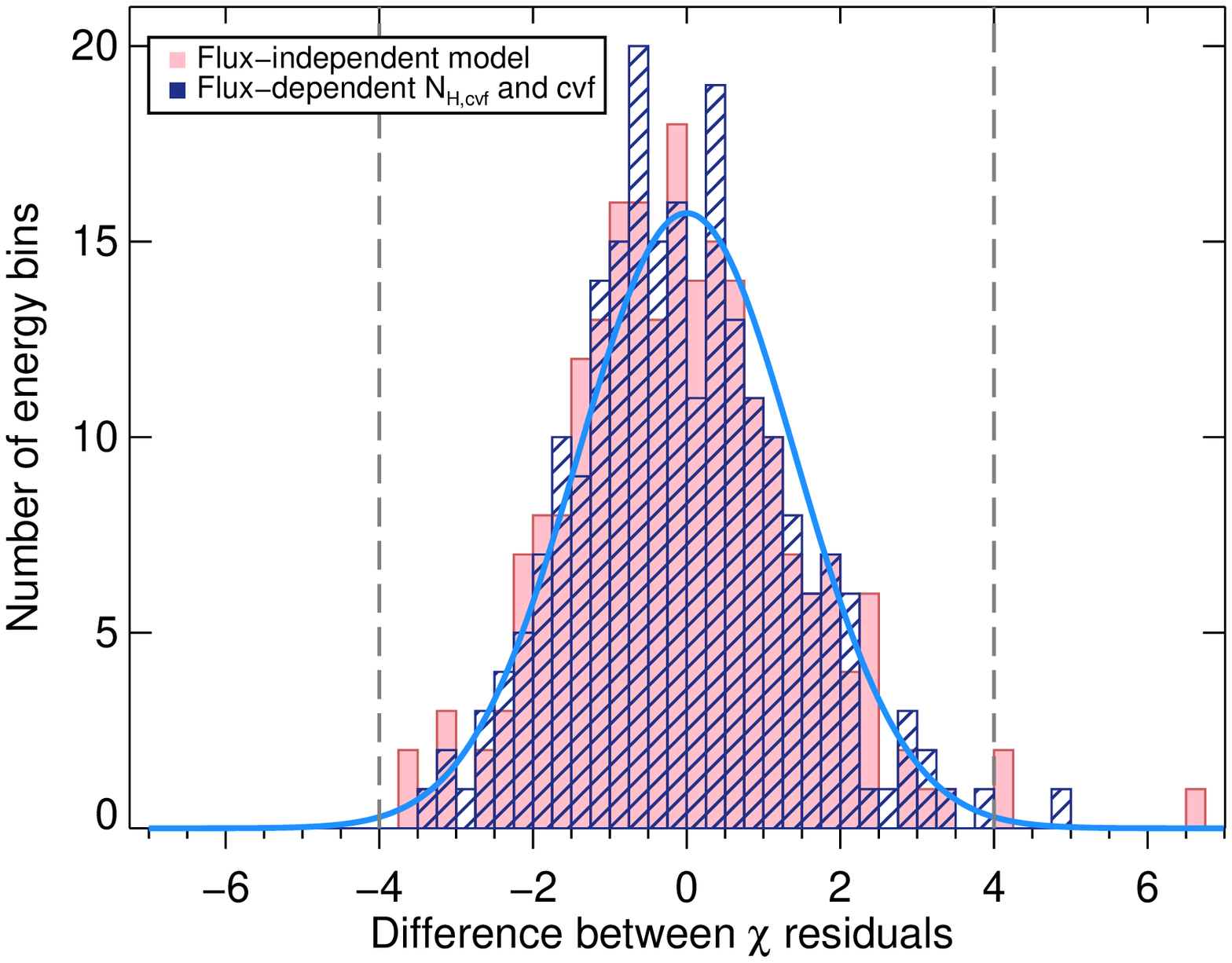}}}
\caption{Distribution of the difference in $\chi$ residuals between the high state and and low state spectra using Model 1 (left) and Model 2 (right).  In each plot, the pink histogram shows the distribution when a flux-independent spectral model is used, while the dark blue striped histogram shows the distribution when the two spectral parameters listed in the legend are allowed to vary between the high state and low state spectra.  The light blue line shows the distribution expected for a good model fit.  The vertical gray dashed lines indicate the limits that only one out of the 223 energy bins should exceed in the case of a good fit.}
\label{fig:chidist}
\end{figure*}

In order to establish whether any statistically significant spectral variations exist between the low and high state spectrum, we fit the low and high state spectra together, allowing only the cross-normalization for each spectrum to vary independently.  Adopting Model 1 (\texttt{constant*tbabs*(bbody+powerlaw*highecut)}) results in $\chi_{\nu}^2$=1.05 for 483 dof, while adopting Model 2 (\texttt{constant*tbabs*pcfabs*powerlaw*highecut}) results in $\chi_{\nu}^2$=1.07 for 483 dof.  The probability that the data for both spectra are drawn from the same model is 22\% for Model 1 and 14\% for Model 2.  Thus, the $\chi_{\nu}^2$ values are statistically consistent with a flux-independent spectral model. \par
However, looking in detail at the $\chi$ residuals to the flux-independent fits shown in Figure \ref{fig:specstates}, it is striking that regardless of whether Model 1 or 2 is adopted, the lowest energy bins are offset by roughly 3$\sigma$ from the model fit, with the models underestimating the high state spectrum and overestimating the low state spectrum below 2~keV.  To assess whether this deviation at soft energies is significant, we rebinned the high state spectrum to match the energy binning of the low state spectrum so as to be able to compare the $\chi$ residuals of the two spectra on a bin-by-bin basis.  We refit the high and low state spectra with Model 1 and Model 2, only allowing the cross-normalization for each spectrum to vary independently.  The largest difference between the $\chi$ residuals of the high state and low state spectra occurs in the lowest energy bin.  For Model 1(2), the lowest energy bin of the high state spectrum is 3.98$\sigma$ (3.32$\sigma$) above the model prediction while the lowest energy bin of the low state spectrum is 2.72$\sigma$ (3.31$\sigma$) below the model prediction, resulting in a net difference of 6.70$\sigma$ (6.63$\sigma$).  If a model is a good fit to the data, we expect the $\chi$ residuals of each spectrum to follow a Gaussian distribution with a standard deviation of 1.0; since the distribution of the difference of two normally distributed variables is also a Gaussian with variance equal to the sum of the variances of the two variables, for a good model fit we expect the difference of the $\chi$ residuals of the high state and low state spectra to follow a Gaussian distribution with a standard deviation of $\sqrt{2}$.  Given that the spectra have 223 energy bins, the difference between the $\chi$ residuals of the two spectra should exceed 4.0$\sigma$ in at most one bin.  As shown by the pink histograms in Figure \ref{fig:chidist}, regardless of whether Model 1 or 2 is used, when the spectral parameters are not allowed to vary between the high and low state, three energy bins have $\chi$ residual differences exceeding 4.0$\sigma$, and the difference of $6.6\sigma-6.7\sigma$ exhibited by the lowest energy bin is much higher than any other bin.\par

\begin{table*}
\centering
\footnotesize
\caption{Variations of Spectral Parameters with Source Brightness}
\begin{threeparttable}
\begin{tabular}{c|ccc|ccc} \hline \hline
\T& \multicolumn{3}{c}{Model 1} \vline & \multicolumn{3}{c}{Model 2} \\
&\multicolumn{3}{c}{\texttt{tbabs*(bbody+powerlaw*highecut)}}\vline&\multicolumn{3}{c}{\texttt{tbabs*pcfabs*powerlaw*highecut}}\\
\B & Both & Low & High & Both & Low & High \\
\hline
\T\B $N_{\mathrm{H}}$ ($10^{22}$ cm$^{-2}$) & & 4.4$^{+0.6}_{-0.4}$ & 3.6$^{+0.7}_{-0.4}$ & 4.0$\pm$1.1 & & \\
\B $kT_{\mathrm{BB}}$ (keV) / $N_{\mathrm{H,partial}}$ ($10^{22}$ cm$^{-2}$) & 1.74$^{+0.06}_{-0.18}$ & & & & 7$\pm2$ & 11$^{+4}_{-3}$ \\
\B BB norm. (10$^{-11}$ erg cm$^{-2}$ s$^{-1}$) / Cov. frac. & & 1.07$\pm0.14$ & 0.85$\pm0.15$ & & 0.9$\pm0.1$ & 0.8$\pm0.1$ \\
\B $\Gamma$ & 0.2$^{+0.5}_{-0.6}$ & & & 1.51$^{+0.10}_{-0.14}$ & &\\
\B PL norm. (10$^{-11}$ erg cm$^{-2}$ s$^{-1}$) & 0.35$^{+0.20}_{-0.13}$ & & & 2.3$^{+0.2}_{-0.3}$ & &\\
\B $E_{\mathrm{cut}}$ (keV) & 12$^{+1}_{-4}$ & & & 7.3$^{+0.7}_{-0.9}$ & &\\
\B $E_{\mathrm{fold}}$ (keV) & 13$^{+5}_{-3}$ & & & 23$^{+5}_{-4}$ & &\\
\B $C_{\mathrm{XMM,pn}}$ & & 1.0 (fixed) & 2.68$^{+0.24}_{-0.22}$ & & 1.0 (fixed) & 2.44$^{+0.12}_{-0.11}$ \\
\B $C_{\mathrm{FPMA}}$ & & 1.07$\pm0.04$ & 2.80$^{+0.24}_{-0.21}$ & & 1.10$\pm0.04$ & 2.63$^{+0.12}_{-0.11}$\\
\B $C_{\mathrm{FPMB}}$ & & 1.10$\pm0.05$ & 2.99$^{+0.25}_{-0.23}$ & & 1.14$\pm0.05$ & 2.81$^{+0.13}_{-0.12}$\\
\B $\chi^2_{\nu}$/dof & 1.01/481 & & & 1.02/481 & &\\
\hline
\end{tabular}
\begin{tablenotes}[flushleft]
\item Notes: The low (high) flux state includes all 100s time intervals of simultaneous \textit{XMM} and \textit{NuSTAR} observations during which the 3--12~keV \textit{NuSTAR} count rate is lower (higher) than $0.5$~counts~s$^{-1}$ (corresponding to a 3--12~keV \textit{XMM-Newton} EPIC pn count rate of $1.3$~counts~s$^{-1}$).  Errors provided are 90\% confidence.  Cross-normalizations between instruments are calculated relative to the \textit{XMM-Newton} EPIC pn instrument low flux state spectrum.  For each spectral model, the low and high flux state spectra are fit together, with the parameters in the ``Both'' column tied to one another.  Abbreviations: BB--blackbody, PL--power law.  The BB normalization is the unabsorbed 0.3--10~keV flux of the BB component.  The PL normalization is the 0.3--10~keV flux of the PL component (in the case of the \texttt{pcfabs} model, it is the flux of the PL component that is subject to local absorption by the partial-covering absorber). 
\end{tablenotes}
\end{threeparttable}
\label{tab:specstates}
\end{table*}

Therefore, we explore whether decoupling one or more spectral parameters between the high state and low state spectral models significantly improves the agreement between the observed and expected distributions of the $\chi$ residual differences.  We find that for both Model 1 and Model 2 a minimum of two spectral parameters must be decoupled between the high state and low state spectral models in order for the $\chi$ residual difference to be $<4\sigma$ in all but 1 of the 223 energy bins.  For Model 1, the two parameters whose decoupling leads to the greatest improvement in $\chi^2_{\nu}$ are the hydrogen column density and the ratio of the blackbody flux to the power-law flux, while for Model 2 they are the covering fraction and the column density of the partial-covering absorber.  The blue striped histograms in Figure \ref{fig:chidist} show the resulting $\chi$ residual difference distribution when these parameters are allowed to vary with source flux in the spectral models. \par
Table \ref{tab:specstates} presents the best-fit results for the flux-dependent models which allow the aforementioned parameters to vary between the high and low flux state.  When adopting Model 1, the high state spectrum is better fit by a lower absorbing column density and a lower ratio of the blackbody flux to the power-law flux than the low state spectrum.  The first trend may suggest that a higher fraction of the local absorbing medium is ionized as the X-ray flux increases, and the second may suggest that the size of the hot spot is larger than the accretion column, resulting in the blackbody emission not being as sensitive to fluctuations in the accretion rate as the power-law emission (Comptonized bremsstrahlung; \citealt{becker07}; \citealt{farinelli16}) from the accretion column.  When adopting Model 2, the high state spectrum is better fit by a partial-covering absorber with a higher column density and a lower covering fraction than the low state spectrum.  This trend is consistent with the accretion rate increasing as the compact object travels through regions where the stellar wind is clumpier.  \par
However, for both Models 1 and 2, the parameters that are allowed to vary between the high state and low state spectral models are still consistent within their 90\% confidence intervals.  Thus, although a flux-dependent spectral model is statistically motivated by the distribution of $\chi$ residual differences shown in Figure \ref{fig:chidist}, the differences in spectral parameters between the high and low flux state are not substantial in magnitude and require better photon statistics to be properly constrained.  These results are consistent with the low level of hardness/softness ratio variations as a function of count rate, and indicate that the fitting results of the flux-averaged spectrum reported in \S\ref{sec:avgspectrum} are not substantially biased by the fact that the \textit{XMM-Newton} and \textit{NuSTAR} observations are not perfectly simultaneous.
%BB model: high flux has lower NH and lower BB/PL > PL responds more readily (accretion column vs. hot spot which may be larger than accretion column)
%PCF model: high flux is higher NH and lower cvf > clumpier structure
%even the parameters that improve fits the most are consistent within 90\% confidence bounds, so while flux-dependent model is statistically motivated, the difference is not substantial 

%there is hint of trend: at higher count rates, softer in XMM (lower NH), harder in NuSTAR (harder/lower gamma) > seen in 4U 2206+54 and  2S 0114+65, slow pulsators (Wang 2013, Farrell et al. 2008; Wang 2011).
%NuSTAR: 1.85e-11 (0.3-10, unabs), NH = 1.4e22
%Chandra: 6e-11 (0.3-10, unabs), NH = 12e22
%XMM: 8e-11(0.3-10, unabs), NH = 3.5e22

\section{Discussion}
\label{sec:discussion}

\subsection{The physical origin of the soft excess}
\label{sec:softexcess}
As discussed in \S\ref{sec:spectral}, the soft excess below 2 keV seen in the spectrum of IGR J18214-1318 can be accounted for either by introducing a blackbody component with properties typical of NS hot spots or a partial-covering absorber associated with the clumpy supergiant wind.  In both models, the column density obscuring the whole binary system is measured to be $N_{\mathrm{H}}\approx4\times10^{22}$ cm$^{-2}$, which can largely be ascribed to the interstellar medium and is consistent with the low column density measured by \citet{rodriguez09}.  The high column density measured by \citet{tomsick08}, well in excess of the ISM value, is comparable to the $N_{\mathrm{H}}$ of the partial-covering absorber.  Thus, the partial-covering model can naturally explain the observed variations in $N_{\mathrm{H}}$ as the result of changes in the density of clumps in the supergiant wind or how deeply embedded the compact object is in the stellar wind at different orbital phases. \par 
However, these observed spectral variations are more difficult to explain using the blackbody model for the soft excess.  Since spectra from the 2008 and 2009 soft X-ray observations of IGR J18214-1318 were fit with simple absorbed power-law models due to their low photon statistics, it is possible that variations in the strength of the blackbody emission could be incorrectly interpreted as $N_{\mathrm{H}}$ variations.  
%If this blackbody emission is indeed associated with hot spots at the NS polar caps, then it should be pulsed due to the NS spin.  Based on our timing analysis, such pulsations would have periods longer than an hour, and thus it is possible that, given their short 5-6 ks exposures, during one of the 2008/2009 observations the hot spot was visible while during the other the hot spot was primarily hidden.  
We made fake 0.3--10 keV spectra with blackbody components of different strengths and fit them with simple absorbed power-law models to determine the effect that the blackbody emission alone can have on the measured $N_{\mathrm{H}}$, but found that it can only account for about 25\% of the measured variations, which span the range of 3--12$\times10^{22}$ cm$^{-2}$).  Although this does not rule out the presence of blackbody emission from this source, it suggests that there are periods of time when significant local obscuration is required to explain the observed $N_{\mathrm{H}}$ measurements.  It is possible that both a partial-covering absorber and blackbody emission are present in this source but the photon statistics of these observations are insufficient to constrain the contribution of both components simultaneously.  \par
In addition to the fact that the partial-covering absorber model provides a more natural explanation for the observed $N_{\mathrm{H}}$ variations, some unusual properties of the blackbody model make the partial-covering model the marginally preferred interpretation for the soft excess.  Although the temperature ($kT\sim1.7$ keV) and size ($R_{\mathrm{BB}}\approx0.3$ km) of the blackbody emission region are typical for NS hot spots, emission from these hot spots is only seen in NS HMXBs with Be or main-sequence donors and $L_X<10^{35}$ erg s$^{-1}$ (\citealt{reig09}; \citealt{bartlett13}).  Furthermore, the flux of the blackbody component of IGR J18214-1318 is 73\%$\pm10$\% of the total 0.3--10 keV flux, which is much higher than the $\sim30$\% hot spot blackbody flux fraction typically seen in other HMXBs (e.g. \citealt{mukherjee05}; \citealt{palombara06}; \citealt{palombara07}; \citealt{palombara09}).  Some Sg HMXBs do exhibit a soft excess, but it is associated with blackbody emission with a lower temperature ($kT\sim0.2$ keV) from a larger area ($R_{\mathrm{BB}}\sim100$~km; \citealt{reig09}).  Such emission is thought to arise from a cloud of diffuse photoionized plasma around the compact object associated with the supergiant wind; the photoionized plasma only absorbs photons at $\gtrsim2$ keV and produces significant emission at $\lesssim1$ keV, resulting in a soft excess (\citealt{hickox04}, \citealt{szostek08}).  However, the soft excess of IGR J18214-1318 has a higher blackbody temperature and smaller radius than is seen in Sg HMXBs.  As a result, we believe that the partial-covering absorber is the most natural way of accounting for the soft excess, even if the blackbody model cannot be definitively dismissed.

%nature of flaring
%BB model: high flux has lower NH and lower BB/PL > PL responds more readily (accretion column vs. hot spot which may be larger than accretion column)
%PCF model: high flux is higher NH and lower cvf > clumpier structure

\subsection{The compact object in IGR J18214-1318}

Since neither pulsations nor cyclotron lines are detected in the \textit{XMM-Newton} and \textit{NuSTAR} data of IGR J18214-1318, we cannot definitively identify the compact object in this HMXB.  Although we cannot rule out that this system hosts a BH, the exponential cutoff to its power-law spectrum with e-folding energy $<25$ keV argues in favor of an NS since BH HMXBs exhibit power-law spectra out to $\gtrsim100$ keV \citep{zdziarski00}.  
Fitting the \textit{NuSTAR} spectrum above 20 keV with a power-law model, we find that $\Gamma=2.5\pm0.2$.  This soft photon index above the cutoff energy is typical for ``normal'' accreting pulsars, whereas anomalous X-ray pulsars (AXPs), which are thought to be magnetars, have $\Gamma=1-2$ above 20 keV \citep{reig12}.  The persistence of the hard X-ray emission from IGR J18214-1318 also disfavors a magnetar origin for this source, since most magnetars (all soft gamma-ray repeaters (SGRs) and many AXPs) exhibit bursting behavior (\citealt{olausen14} and references therein).  Thus, the compact object in IGR J18214-1318 is most likely an accreting NS with a typical magnetic field strength (10$^{12}-10^{13}$ G).  The lack of detected pulsations with periods $\lesssim1$ hrr can be explained by the geometry of the system (e.g., the nearly perfect alignment of the magnetic and rotational axes of the NS, the NS beam being narrow and not pointing toward Earth, or the NS beam being broad enough so as to wash out spin modulations) or a spin period $>1-2$ hr, longer than is typically seen in NS HMXBs.  

\subsection{Comparison to other HXMBs}

\begin{table*}
\centering
\footnotesize
\caption{Other non-pulsating and long pulsation period HMXBs}
\begin{threeparttable}
\begin{tabular}{cccccp{0.7in}cccp{1.5in}} \hline \hline
\T Source name & $P_{\mathrm{spin}}$ & $P_{\mathrm{orbit}}$ & Donor & Distance & $L_X$ (10$^{35}$ & $N_{\mathrm{H}}$ & $\Gamma$ & $E_{\mathrm{fold}}$ & \hspace{0.5in}References \\
 & (ks) & (days) & type & (kpc) & erg cm$^{-2}$ s$^{-1}$) & ($10^{22}$ cm$^{-2}$) & & (keV) &\\
\B (1) & (2) & (3) & (4) & (5) & (6) & (7) & (8) & (9) & \hspace{0.7in}(10) \\
\hline
\T\B Non-pulsating HMXBs &&&&&&&&& \\
\hline
\T 4U 1700-377 & $>50$ & 3.41 & O6.5 Iaf & 1.9 & 0.4-40 \hspace{0.2in} (0.5--12 keV) & 5--80 & 0.6--1.9 & 10--24 & \citet{gottwald86}; \citet{heap92}; \citet{reynolds99}; \citet{ankay01}; \citet{boroson03}; \citet{vandermeer05} \\
IGR J08262-3736 & $>5$ & -- & OB V & 6.1 & 0.2--0.4 \hspace{0.2in} (2--10 keV)& 0.6--1.4 & 1.3--1.8 & $>70$ & \citet{masetti10}; \citet{bozzo12}\\
IGR J16207-5129 & $>7$ & -- & B1 Ia & 6.1 & 0.2--5.0 \hspace{0.2in} (0.5--10 keV) & 15--17 &  0.9--1.3 & 14--28 & \citet{nespoli08}; \citet{tomsick09}; \citet{ibarra07}; \citet{bodaghee10}\\
IGR J16318-4848 & $>10$ & -- & sgB[e] & 4.8 & 1.7 (20-50 keV) 66 (5-60 keV) & 100--200 & 0.5--1.5 & 17--60 & \citet{walter03}; \citet{filliatre04}; \citet{barragan09}\\
\B IGR J19140+0951 & $>2$ & 13.55 & B0.5 Ia & 5.0 & 0.3--15 \hspace{0.25in} (2--20 keV) & <1-20 & 1.0--2.0 & 6--10 & \citet{hannikainen07}; \citet{prat08}\\
\hline
\T\B Long $P_{\mathrm{spin}}$ HMXBs &&&&&&&&& \\
\hline
\T 2S 0114+650 & 9.7 & 11.59 & B1 Ia & 4.5 & 1.5 (2-10 keV) 4.3 (3-20 keV) & 2--5 & 0.8--1.2 & 13--18 & \citet{reig96}; \citet{corbet99}; \citet{farrell08}\\
\B 4U 2206+54 & 5.56 & 9.56 & O9.5 V & 2.6 & 0.35--3.5 \hspace{0.2in} (1--12 keV) & 0.9--4.7 & 0.9--1.8 & 10--29 & \citet{blay05}; \citet{ribo06}; \citet{reig09} \\
\hline
\end{tabular}
\begin{tablenotes}[flushleft]
\item Notes: (2) Pulsation period.  For non-pulsating HMXBs, the value provided is the longest period that has been ruled out by periodicity searches.  It is possible that these sources have pulsation periods longer than these values.
\item (3) Orbital period of the binary (for sources for which it has been measured).
\item (4) Spectral type of the donor star.
\item (5) Distance estimate.  Typical uncertainty is $\pm2$ kpc.
\item (6) Range of observed X-ray luminosities for the source.  The energy band is written in parentheses.
\item (7-9) Range of observed spectral parameters based on different observations of the source.
\end{tablenotes}
\end{threeparttable}
\label{tab:nonpulsating}
\end{table*}

IGR J18214-1318 is one of only a handful of HMXBs that may host an NS but for which sensitive periodicity searches have not discovered pulsations in the spin period range of $0.1-2000$ s, which is typical for NSs in HMXBs.  We refer to these HMXBs as non-pulsating, although it cannot be ruled out that they have pulsations with longer periods than are probed by current observations.  Table \ref{tab:nonpulsating} provides an overview of the properties of non-pulsating HMXBs, as well as HMXBs with long pulsation periods ($P_{\mathrm{spin}}>$1 hr) as a point of comparison.  \par
As can be seen in Table \ref{tab:nonpulsating}, all but one of the non-pulsating HMXBs exhibit exponential cutoffs to their power-law spectra with e-folding energies $\lesssim30$~keV, suggesting that they likely harbor NSs.  The one exception is IGR J08262-3736, for which no exponential cutoff is measured below 70 keV \citep{bozzo12}.  The lack of an exponential cutoff in the spectrum of IGR J08262-3736 makes it the most plausible BH candidate of all the non-pulsating HMXBs.  This HMXB also differs from all the other non-pulsating HMXBs, including IGR J18214-1318, in that its donor star is a main-sequence OB star \citep{masetti10}. \par
%One way in which IGR J08262-3736 resembles IGR J18214-1318 is in exhibiting a soft excess which can be accounted for either by a partial absorber or a blackbody component consistent with emission from NS hot spots ($kT_{\mathrm{BB}}\approx1.2$ keV, $R_{\mathrm{BB}}\approx0.2$ km; \citealt{bozzo12}).  While we cannot rule out either model for either of these HMXBs, blackbody emission with such a high temperature has never been seen before in a source which, like IGR J18214-1318, is a Sg HMXB, but has been observed in sources which, like IGR 08262-3736, have a main-sequence companion. }  \par
Of the four non-pulsating Sg HMXBs, IGR J16207-5129 most resembles IGR J18214-1318.  In addition to having a similar X-ray luminosity, photon index, and exponential cutoff energy, this source also exhibits a soft excess, which \citet{tomsick09} argue most likely results from partial obscuration by the stellar wind with $N_{\mathrm{H}}\approx10^{23}$ cm$^{-2}$.  The PDS of IGR J16207-5129, like the PDS of IGR J18214-1318, shows significant red noise below 0.01 Hz, which is well-fit by a simple power law; the slope of the red noise in IGR J16207-5129 is steeper, having $\alpha=1.76\pm0.05$ \citep{tomsick09}, but still falling within the range that is typical for frequencies above the pulsation frequency, suggesting, but not proving, that the NS in this HMXB may have a period longer than the $\sim2$ hr limit probed by the data.\par
Another non-pulsating Sg HMXB that is similar to IGR J18214-1318 is IGR J19140+0951, except for the fact that IGR J19140+0951 exhibits a soft excess modeled by a lower temperature ($kT=0.3$ keV) blackbody, which is attributed to a shock formed between the ionized gas around the NS and the stellar wind \citep{prat08}.  The $N_{\mathrm{H}}$ variations measured in the spectrum of IGR J19140+0951 are related to its orbital phase and are even larger than those observed for IGR J18214-1318. \par
% Additional observations of IGR J18214-1318 are required to determine its orbital period and determine whether any of its spectral properties vary with orbital phase.}\par
Apart from having similar photon indices and exponential cutoff energies indicative of NS HMXBs, the other two non-pulsating Sg HMXBs have more extreme properties than IGR J18214-1318.  IGR J16318-4848 is the HMXB with the highest measured local column density ($N_{\mathrm{H}}\sim10^{24}$~cm$^{-2}$; \citealt{walter03}), which is due to its sgB[e] donor star \citep{filliatre04}.  The relatively slow 400~km~s$^{-1}$ wind of this donor star helps maintain a high mass accretion rate onto the compact object, resulting in a slightly higher X-ray luminosity ($L_X\sim10^{36}$~erg~s$^{-1}$) than in other Sg HMXBs \citep{barragan09}.  4U 1700-377 displays the highest levels of variability of any of the non-pulsating Sg HMXBs; its flux, even at hard X-ray energies, varies by factors $>$100, and the pattern of spectral variations as a function of luminosity provides further evidence that 4U 1700-377 likely hosts an NS \citep{seifina16}.  However, the mass of the compact object in this system has been measured to be 2.44$\pm$0.27 $M_{\odot}$ \citep{clark02}, placing it among the most massive NSs observed \citep{ozel16}.  For 4U 1700-377 and IGR J16318-4848, the existence of pulsations is constrained over a wider range of periods than for other sources, excluding all periods $<50$ ks and $<10$ ks, respectively.  \par
It is possible that some of the non-pulsating HMXBs, especially those for which periodicity searches do not extend beyond 1--2 hr, possess slowly spinning NSs.  Currently, two HMXBs, 4U 2206+54 and 2S 0114+650, are known to have pulsation periods longer than an hour (\citealt{reig09}; \citealt{corbet99}).  The photon indices and cutoff energies of both sources are similar to IGR J18214-1318 (\citealt{farrell08}; \citealt{reig09}), although 2S 0114+650 is a closer analogue since it has a supergiant donor \citep{reig96}.  \citet{li99} proposed that the slow spin of the NS in 2S 0114+650 indicates that it was born as a magnetar with $B\gtrsim10^{14}$ G, was slowed down efficiently by the propeller effect before its magnetic field significantly decayed to its current expected value of $\sim10^{12}$ G.  In the case of 4U 2206+54, magnetorotational models, which can account for the NS's spin and spin-down rate, require magnetic fields strengths between $5\times10^{13}$ and $3\times10^{15}$ G \citep{ikhsanov10}; thus, it is possible that 4U 2206+54 currently contains a magnetar and would evolve into a system like 2S 0114+650.  Neither source exhibits clear cyclotron absorption features, although low-significance detections of such features have been claimed by some authors, which would indicate that both HMXBs host NSs with typical strength magnetic fields (\citealt{bonning05}; \citealt{torrejon04}).  \par
Neutron stars with long spin periods have also been discovered in some symbiotic X-ray binaries (SyXBs).  The two SyXBs with the longest spin periods, IGR J16358-4726  and 4U 1954+319, have spin periods of  $\sim$1.6 hr and $\sim$5.3 hr, respectively (\citealt{kouveliotou03}; \citealt{patel04}; \citealt{corbet06}; \citealt{marcu11}).  Both of these sources have spectra that are very similar to those of long spin-period HMXBs and IGR J18214-1318 (\citealt{patel07}; \citealt{enoto14}), but they have giant M-type stellar companions rather than high-mass donors.  
%While SyXBs with slowly spinning NSs share many properties in common with long-period HMXBs and IGR J18214-1318, their late-type stellar counterparts indicate a distinct evolutionary origin.  
Although it has been suggested that the long spin-period SyXBs may host magnetars or magnetar descendants just like long spin-period HMXBs (\citealt{patel07}; \citealt{enoto14}), models that assume quasi-spherical wind accretion for SyXBs rather than disk accretion do not require magnetar-strength fields to explain their timing properties \citep{lu12}.   \par
In summary, in addition to IGR J18214-1318, there are five non-pulsating HMXBs, four of which have exponential cutoff energies $<30$ keV, suggesting they most likely harbor NSs.  Like IGR J18214-1318, all four of these HMXBs have supergiant donor stars and resemble the Sg HMXB 2S 0114+650, which hosts an NS with a 2.7 hr pulsation period thought to have been born as a magnetar.  Population synthesis models predict that 8-9\% of all NSs are born as magnetars, and that only $\sim2$\% of NSs in binaries are magnetars; these models predict that an even a smaller percentage of magnetars would be part of an X-ray binary because many of them are produced from the secondary rather than the primary \citep{popov06}.  About 100 HMXBs have been discovered in the Galaxy, and only about 60 of them are known to host pulsars \citep{bird16}, so with the discovery of 2S 0114+650 and 4U 2206+54, both of which may host magnetars (or former magnetars), the observed number of magnetar HMXBs already agrees with theoretical expectations.  Thus, if future observations reveal that several of the ``non-pulsating'' HMXBs actually host long-period pulsars, it could imply that either current models of magnetar origins or models of the spin evolution of NSs in binaries need to be revised.  
		
\section{Conclusions}

Timing analysis of the \textit{XMM-Newton} and \textit{NuSTAR} observations of IGR J18214-1318 shows that this HMXB has strong levels of aperiodic variability but no pulsations with periods shorter than an hour.  Joint fitting of the \textit{XMM-Newton} and \textit{NuSTAR} spectra reveals the presence of an exponential cutoff with e-folding energy $<25$ keV.  Thus, although we cannot definitively identify the nature of the compact object in this system, the spectral cutoff energy is a strong indication that it is an NS.  A soft excess is also detected in the spectrum of IGR J18214-1318, which may result from the inhomogeneous supergiant wind partially obscuring the X-ray source or blackbody emission with $kT=1.74^{+0.04}_{-0.05}$ keV and $R_{BB}\approx0.3$ km, which are values typical for emission from hot spots on an NS surface. The partial-covering absorption model is marginally preferred since it can explain the large variations of the total column density measured by different observations; furthermore, the blackbody temperature and the ratio of the blackbody flux to the total X-ray flux measured for IGR J18214-1318 are significantly higher than for other Sg HMXBs, in which blackbody emission seems to arise from the supergiant wind rather than hot spots. \par
%A soft excess is also detected in the spectrum of IGR J18214-1318, which we argue most likely results from partial covering absorption from the supergiant wind; the column density associated with the stellar wind is $\sim10^{23}$ cm$^{-2}$, adding this source to the group of highly obscured Sg HMXBs discovered by \textit{INTEGRAL}.  \par
This study demonstrates the usefulness of the combination of \textit{XMM-Newton} and \textit{NuSTAR} observations to identify the likely nature of compact objects in HMXBs.  Similar observations of other HMXBs will help place better constraints on the BH/NS ratio in HMXB populations.  We know of four other HMXBs which, like IGR J18214-1318, have supergiant donor stars, cutoff energies $\lesssim30$ keV suggesting they harbor NSs, but no detected pulsations despite sensitive timing observations.  These sources also resemble 2S 0114+950, an Sg HMXB with 2.7 hr pulsations thought to host a former magnetar.  Determining how many of the non-pulsating HMXBs in fact harbor long-period pulsars could shed light on the origins of magnetars and their possible connection to long-period pulsars in HMXBs.  

\acknowledgments

We thank the anonymous referee for suggestions that improved the clarity and thoroughness of this paper.  This work made use of data from the \textit{NuSTAR} mission, a project led by the California Institute of Technology, managed by the Jet Propulsion Laboratory, and funded by the National Aeronautics and Space Administration.  We thank the \textit{NuSTAR} Operations, Software and Calibration teams for support with the execution and analysis of these observations. This research has made use of the \textit{NuSTAR} Data Analysis Software (NuSTARDAS) jointly developed by the ASI Science Data Center (ASDC, Italy) and
the California Institute of Technology (USA).  In addition, F.M.F. acknowledges support from the National Science Foundation Graduate Research Fellowship.  F.M.F. and J.A.T. acknowledge partial support from NASA under {\em XMM} Guest Observer grant NNX15AG31G.  R.K. acknowledges support from the Russian Science Foundation (grant 14-12-01315).  L.N. wishes to acknowledge financial support by ASI/INAF grant I/037/12/0.

% BIBLIOGRAPHY
\bibliographystyle{aasjournal}
\bibliography{refs}

\end{document}